\documentclass[lettersize,journal]{IEEEtran}

\usepackage{mathrsfs}
\usepackage{latexsym,amssymb,amsmath,alltt}
\usepackage {stfloats}
\usepackage {graphicx}
\usepackage {subfigure}
\usepackage {url}
\usepackage {cite}
\usepackage{subfigure}
\usepackage{colortbl}
\usepackage{graphicx}  % Written by David Carlisle and Sebastian Rahtz
\usepackage{url}       % Written by Donald Arseneau
\usepackage{epstopdf}
\usepackage{amsmath}   % From the American Mathematical Society
\usepackage{cite}
\allowdisplaybreaks[4]
\usepackage{amsfonts,amssymb}

\usepackage{stfloats}
\usepackage{algorithm}
\usepackage{algorithmic}
\usepackage{cases}
\usepackage{algorithm}
\usepackage{multirow}
\usepackage{algorithmic}
\usepackage{bm}
\newtheorem{theorem}{{Theorem}}
\newtheorem{lemma}{{Lemma}}

\makeatletter
\newcommand{\vvast}{\bBigg@{3.0}}
\newcommand{\vast}{\bBigg@{4}}
\newcommand{\Vast}{\bBigg@{4.5}}
\newcommand{\VVast}{\bBigg@{5}}
\newcommand{\VVVast}{\bBigg@{5.5}}


\newcommand{\ls}[1]
{\dimen0=\fontdimen6\the\font
	\lineskip=#1\dimen0
	\advance\lineskip.5\fontdimen5\the\font
	\advance\lineskip-\dimen0
	\lineskiplimit=.9\lineskip
	\baselineskip=\lineskip
	\advance\baselineskip\dimen0
	\normallineskip\lineskip
	\normallineskiplimit\lineskiplimit
	\normalbaselineskip\baselineskip
	\ignorespaces
}

\graphicspath{{Visio-File-0731-final/}{Simulation-1213/}}

\begin{document}

%\date{March 17, 2019}

\title{FBC-Enhanced $\epsilon$-Effective Capacity Optimization for NOMA}
	
	\author{Jingqing Wang, Wenchi Cheng, and Wei Zhang\vspace{-2pt}
		\\
		%Here I still need to add corresponding author's information to make the first paper filled.
		\thanks{\ls{.5}
			\setlength{\parindent}{2em}
			
			Jingqing Wang and Wenchi Cheng are with Xidian University, Xi'an, 710071, China (e-mails: jqwangxd@xidian.edu.cn; wccheng@xidian.edu.cn).
			
			Wei Zhang is with the School of Electrical Engineering and Telecommunications, The University of New South Wales, Sydney, Australia (e-mail: w.zhang@unsw.edu.au).
			%Jiangzhou Wang is with the School of Engineering and Digital Arts, University of Kent, Canterbury CT2 7NZ, U.K. (e-mail: j.z.wang@kent.ac.uk).
		}
		%\thanks{Digital Object Identifier 0000000000000000}
		%\vspace{-55pt}
	}
	%	\date{May 1, 2023}
	
	%\markboth{TO BE SUBMITTED TO IEEE Wireless Communications}{TO BE SUBMITTED TO IEEE Wireless Communications}

\maketitle

%\vspace{-34pt}
\begin{abstract}	
The advent of massive ultra-reliable and low-latency communications (mURLLC) has introduced a critical class of time- and reliability-sensitive services within next-generation wireless networks. This shift has attracted significant research attention, driven by the need to meet stringent quality-of-service (QoS) requirements.
In this context, non-orthogonal multiple access (NOMA) systems have emerged as a promising solution to enhance mURLLC performance by providing substantial enhancements in both spectral efficiency and massive connectivity, particularly through the development of finite blocklength coding (FBC) techniques.
Nevertheless, owing to the dynamic nature of wireless network environments and the complex architecture of FBC-enhanced NOMA systems, the research on the efficient design of optimizing the system performance for maximizing system capacity while guaranteeing the tail distributions in terms of new statistical QoS constraints for delay and error-rate is still in its infancy.  
In an effort to address these challenges, we put forth the formulation and solution of $\epsilon$-effective capacity problems tailored for uplink FBC-enhanced NOMA systems, specifically catering to ensure statistical delay and error-rate bounded QoS requirements.
In particular, we establish uplink two-user FBC-enhanced NOMA system models by applying the hybrid successive interference cancellation (SIC). 
We also develop the concept of the $\epsilon$-effective capacity and propose the optimal power allocation policies to maximize the $\epsilon$-effective capacity and $\epsilon$-effective energy efficiency while upper-bounding both delay and error-rate. 
We conduct a set of simulations to validate and evaluate our developed optimization schemes over FBC-enhanced NOMA systems.
\end{abstract}


%\vspace{-1.7cm}
\section{Introduction}\label{sec:intro}
Amid the global deployment of 5G wireless networks, researchers are actively formulating visions for next generation mobile wireless networks~\cite{YEH202382}. 
%These endeavors aim to facilitate the realization of exceptionally advanced wireless applications, characterized by increasingly stringent and diverse technical performance requirements, including but not limited to massive connectivity, stringent end-to-end delay, ultra-high reliability, and unparalleled data rates. 
A pivotal emphasis in the development of 6G revolves around devising systems capable of enabling \textit{Ultra-Reliable Low-Latency Communications} (mURLLC)~\cite{10175044,10175166}, which necessitates the provision of diverse quality-of-service (QoS) guarantees to ensure the timely and reliable data deliveries. 
Architectures, schemes, and algorithms for statistical QoS theory~\cite{cheng2016,10609803}, considering the effective capacity function, have been introduced and demonstrated to be potent methods for jointly characterizing delay-bounded QoS constraints across distinct wireless links.
In the context of real-time multimedia applications, it is imperative to develop potential QoS control strategies that optimizes the system capacity while ensuring the stringent mURLLC requirements.	

However, a significant challenge arises in navigating the critical obstacle posed by expansive connectivity, particularly in the context of supporting QoS-driven mURLLC services within the 6G landscape.
The development of non-orthogonal multiple access (NOMA)~\cite{9693417,9036072} has emerged to provide substantial enhancements in both spectral efficiency and massive connectivity performance, while guaranteeing stringent QoS benchmarks requisite.
%NOMA-based access schemes can significantly improve the spectrum efficiency and potentially solve the massive connectivity issues as compared with conventional orthogonal multiple access (OMA) schemes. 
%NOMA-based mobile users are required to perform full decoding of all interference when receiving messages, employing \textit{successive interference cancellation} (SIC) as outlined in~\cite{z2017}. This approach notably escalates the computational complexity associated with signal processing. 
%The authors of~\cite{9693417} have explored the evolution of Next Generation Multiple Access (NGMA) with a particular focus on NOMA.
Numerous research endeavors are dedicated to the development of optimization frameworks for NOMA systems.
Specifically, to minimize task computation delay, the authors of~\cite{9741761} have studied the single-user multi-edge-server for Multi-access edge computing system based on downlink NOMA.
In addition, the authors of~\cite{9547827} have proposed the unmanned aerial vehicles (UAVs)-assisted full-duplex (FD) NOMA system for improving the overall sum-rate throughput by joint user clustering, optimal UAV placement and power allocation. 
The authors of~\cite{9520776} have proposed an intelligent reflecting surface (IRS)-assisted NOMA scheme to achieve secure communication via artificial jamming.

Nevertheless, existing studies have primarily concentrated on developing potential optimization strategies for NOMA systems within the asymptotic regime to increase the system ergodic capacity, neglecting the scenarios involving the finite blocklength.
In contrast to the conventional orthogonal multiple access (OMA) architecture, the NOMA framework, particularly in the context of short packet communications, was devised with the primary aim of significantly enhancing spectral efficiency and mitigating communication delays~\cite{9151196,8936382}. 
The authors of~\cite{9499078} have developed the a semi-grant-free NOMA scheme that admitting the grant-free users is completely transparent to the grant-based users to guarantee the quality-of-service experience.

To fulfill the delay requisites of mURLLC, the utilization of small-packet communications, specifically finite blocklength coding (FBC)~\cite{yury2010,Yp2011,G2013,9635675,10177877}, assumes a pivotal role in assessing and regulating the reliability metric for enhancing the performance of NOMA systems. 
Traditionally, Shannon's second theorem~\cite{cover1991} posits that for any given error probability $\epsilon$ and information rate, a code can be devised such that, given a sufficiently large coding block length $n$, the maximum probability of block error does not exceed $\epsilon$. 
%The researchers in~\cite{yury2010} have presented methodologies for characterizing the FBC-based maximum achievable coding rate through the additive white Gaussian noise (AWGN) channels.
As demonstrated in~\cite{B2015}, ensuring error-rate bounded constraints necessitates a short coding blocklength. 
%The authors of~\cite{G2013} have conducted an analysis of the throughput in FBC-based cognitive radio systems, considering queuing constraints. 
In~\cite{Durisi2016}, the authors devised approximations for the maximal data transmission rate, considering both the decoding error probability and channel capacity through applying FBC. 
Therefore, considering the FBC-enhanced NOMA systems, the authors of~\cite{9942348} have investigated the performance of NOMA systems with hybrid long- and short-packet communications.
Additionally, the authors of~\cite{10041974} have analyzed the performance of an intelligent reflecting surface (IRS)-aided short-packet NOMA system over Nakagami-$m$ fading channels.
The authors of~\cite{9663542} have performed an unified analysis of FBC-NOMA system with QoS-based  successive interference cancellation (SIC) detecting order. 
In~\cite{10292880}, the authors have explored the max-min fairness for uplink NOMA in the finite blocklength regime.

More specifically, it becomes essential to effectively control and manage the system performance by designing QoS-driven resource allocation strategies for delay and error-sensitive wireless applications while guaranteeing mURLLC in FBC-enhanced NOMA communications.
However, the cross-layer optimization problems typically lack convexity or concavity concerning transmit power or bandwidth. This inherent complexity poses challenges in solving the optimization problems for FBC-enhanced NOMA systems.
%Accordingly, to ensure mURLLC services, the authors in~\cite{Chen2017} have reviewed the applications, design challenges, and potential approaches to reduce the end-to-end latency.
Researchers, such as those in~\cite{10184225}, have surveyed the state-of-the-art work in the literature of NOMA focusing on analyzing the error-rate performances.
Furthermore, the work by the authors in~\cite{9964377} has investigated the system performance of UAV-aided FBC-enhanced NOMA systems under imperfect channel state information (CSI) and SIC.
Meanwhile, the researchers of~\cite{10175044} have proposed a joint sub-channel allocation and power control method to support NOMA-mURLLC wireless networks. 
%Meanwhile, the authors of~\cite{9036072} have investigated the potential reduction in queueing delay by employing uplink NOMA compared to OMA, especially in scenarios where the system necessitates communication with very low latency and high reliability.

Despite the diligent efforts from both academia and industry for guaranteeing mURLLC services, the majority of prior investigations have primarily concentrated on designing the FBC-enhanced algorithms/architectures/frameworks without considering both delay and reliability aspects in terms of statistical QoS.
It becomes crucial to design and model FBC-enhanced NOMA architectures with due consideration for stringent and diverse  QoS guarantees related to both delay and error rates. 
This entails the introduction of a new $\epsilon$-effective capacity function and the subsequent formulation and resolution of FBC-based $\epsilon$-effective capacity maximization problems.
However, due to the lack of comprehensive grasp of QoS-driven fundamental performance-limits and modeling analyses over FBC-enhanced NOMA systems, how to efficiently identify and define novel FBC-based $\epsilon$-effective capacity function bounded by both \textit{delay} and \textit{error-rate} is still a challenging task.
Moreover, the intricate task of skillfully formulating the corresponding optimization problems, particularly when integrating FBC techniques, gives rise to numerous challenges. 
Subsequently, there remain significant hurdles in resolving the associated QoS-driven optimization problems for uplink FBC-enhanced NOMA systems while simultaneously upper-bounding both delay and error rates.
%The stringent ad diverse QoS requirements of mURLLC requires a new paradigm for resource allocations in view of the limitations of existing NOMA based schemes.

To address the above challenges, in this paper we propose to formulate and solve of the FBC-based $\epsilon$-effective capacity maximization problem by devising optimal power allocation policies for FBC-enhanced NOMA systems, all while adhering to statistical constraints on delay and error rates within the framework of QoS provisioning.
Particularly, we commence by defining system architecture models for uplink two-user FBC-enhanced NOMA in the context of wireless communications, with a focus on FBC-based channel coding rates.
In the context of statistical QoS demands, characterized by constraints on both delay and reliability, we articulate and address the $\epsilon$-effective capacity maximization problems associated with our proposed uplink two-user FBC-enhanced NOMA schemes. Additionally, we deduce the optimal average power allocation policy for maximizing $\epsilon$-effective energy efficiency with unknown CSI.
Subsequently, we conduct a series of simulations to verify, assess, and evaluate the efficacy of the developed optimal power allocation policies in the two-user FBC-enhanced NOMA schemes.

The subsequent sections of this paper are structured as follows. 
In Section~\ref{sec:sys}, we build an uplink two-user NOMA system model that integrates FBC. 
Following this, Section~\ref{sec:qos1} is dedicated to the formulation and resolution of the $\epsilon$-effective capacity optimization problem, taking into consideration a constrained decoding error probability within the finite blocklength regime.
In Section~\ref{sec:qos2}, we proceed to formulate and address the $\epsilon$-effective energy efficiency optimization problem.
Section~\ref{sec:results} is devoted to the evaluation of the performance of our proposed two-user FBC-enhanced NOMA system. The paper concludes with Section~\ref{sec:conclusion}.

\section{The FBC-Enhanced  NOMA Based System Model}\label{sec:sys}

\begin{figure}[t]
	\centering
	\includegraphics[scale=0.22]{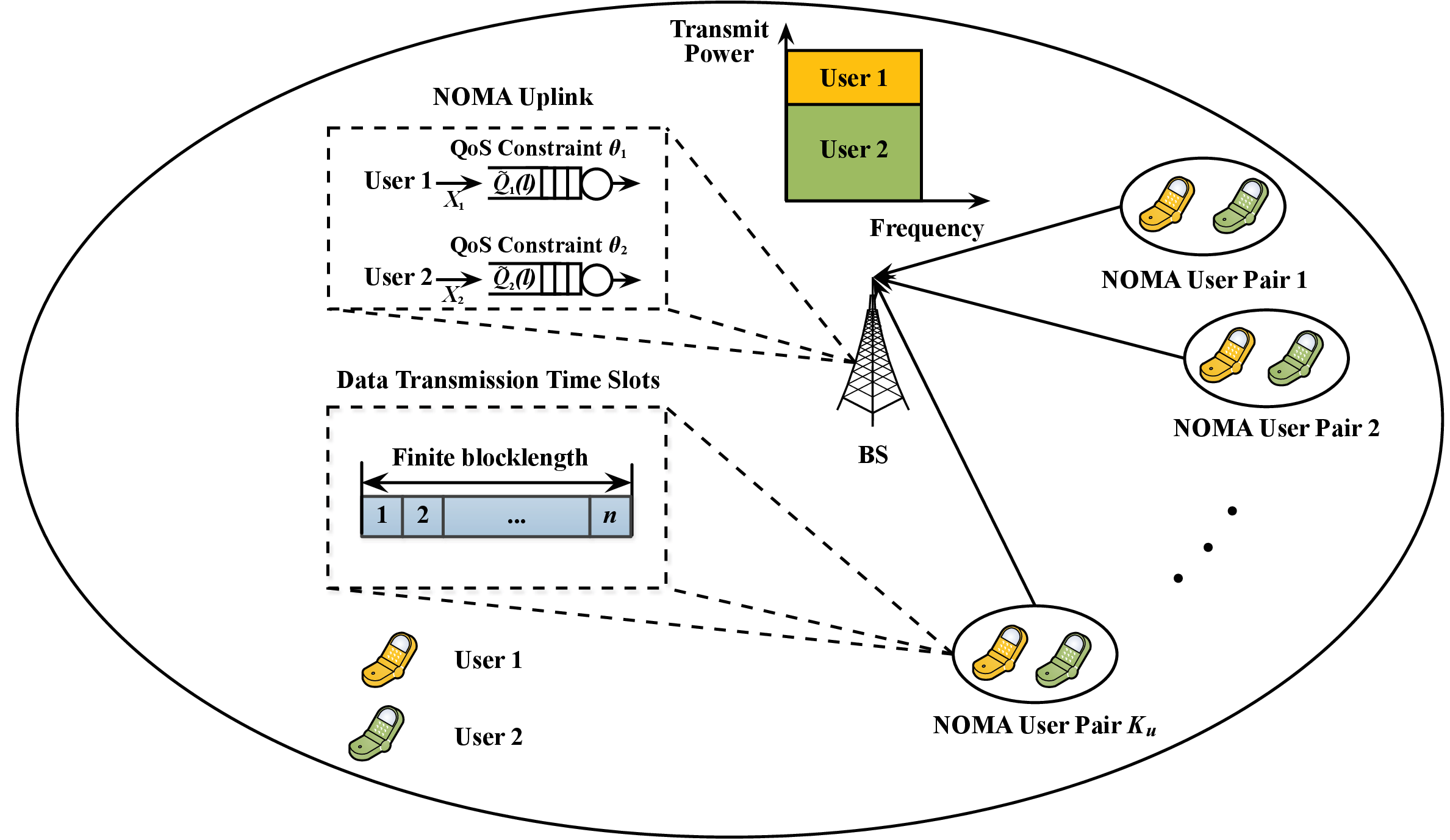}
	\caption{A system architecture model for uplink two-user FBC-enhanced NOMA system, where $\widetilde{Q}_{i}(l)$ denotes the queue length process for mobile user $i$ $(i=1,2)$ and $\theta_{i}$ denotes the QoS exponent for mobile user $i$ $(i=1,2)$.}
	\label{fig:1}
\end{figure}

Consider a two-user FBC-enhanced NOMA system model, where there is one base station (BS) and ${\cal K}_{\text{u}}=\{1,2,\dots, k_{u},$ $\dots, K_{\text{u}}\}$  $(K_{\text{u}}=|{\cal K}_{\text{u}}|)$  denotes the set of NOMA two-mobile-user pairs with each element consisting of two different mobile users among which mobile user 1 is relatively near from the BS (thus being allocated with relatively less transmit power) while mobile user 2 is relatively far from the BS (thus being allocated with relatively more transmit power), as depicted in Fig.~\ref{fig:1}. 
%Uplink mobile users upload their messages with short packets to the BS. 
%Without loss of generality, we assume that there are two mobile users (U$_{1}$ and U$_{2}$) in each NOMA pair that share the same bandwidth resources using different codebooks. 
The wireless channel is presumed to exhibit block-fading characteristics, wherein it remains constant throughout the duration of one block comprising $n$ channel uses.
This two-user scenario can be readily extended to an $N$-user scenario.

\subsection{Two-User Uplink FBC-Enhanced  NOMA Based System Models With Hybrid SIC Decoding}

%\begin{equation}
%n^{(i)}=\sum_{k=1}^{K}n^{(i)}.
%\end{equation}
%We assume that the codeword blocklength for mobile user 1 and mobile user 2 are the same, i.e., $n^{(1)}=n^{(2)}=n$. Without loss of generality, we use $n$ instead of $n^{(1)}$ and $n^{(2)}$ in the following sections.
%Define the transmitted signal vector $\bm{X}^{(i)}=[\bm{X}_{1}^{(i)},\dots, \bm{X}_{K}^{(i)}]$ and the received signal vector $\bm{Y}^{(i)}=[\bm{Y}_{1}^{(i)},\dots, \bm{Y}_{K}^{(i)}]$ for mobile user $i$.
%We assume that each codeword $\bm{x}_{k}$ satisfies the following  condition:
%\begin{equation}
%\|\bm{x}_{k}\|^{2}= {\cal P}_{i},
%\end{equation}
%We define ${\cal P}_{i}^{(i)}$ as the power of transmitted signal $X_{k}^{(i)}$ associated with $k$th subchannel for mobile user $i$. %the codeword $\bm{x}_{m}$ ($m=1,\dots, M$).
%As a result, define the transmit power vector ${\cal P}_{2}^{(i)}=\left[{\cal P}_{1}^{(i)},\dots, {\cal P}_{i}^{(i)}\right]$ at mobile user $i$, satisfying the following average power constraint:
%\begin{equation}
%\frac{1}{K}\sum_{k=1}^{K}\frac{n^{(i)}}{n}{\cal P}_{i}^{(i)}\leq \overline{{\cal P}}^{(i)},
%\end{equation}
%where $\overline{{\cal P}}^{(i)}$ $(i=1,2)$ represent the average transmit power at the BS to mobile user $i$. 

In a considered NOMA pair, we define $X_{i,l}$ as the signal intended for the uplink data transmissions between mobile user $i$ and the BS in time slot $l$, where $l = 1,\dots,n$. 
Define the uplink transmit data sequence as $\bm{X}_{i}=(X_{i,1},\dots, X_{i,n})$ for mobile user $i$ with $i\in\{1,2\}$.
We can write the received signal, denoted by $\bm{Y}$, at the BS for the uplink two-user FBC-enhanced NOMA system as follows:
\begin{align}\label{equation01}
\bm{Y}={\cal P}_{1}\bm{H}_{1}\bm{X}_{1}+{\cal P}_{2}\bm{H}_{2}\bm{X}_{2}+\bm{W},
\end{align}
where ${\cal P}_{i}$ $(i=1,2)$ denotes the transmit power at mobile user $i$, $\bm{X}_{i}$ $i\in\{1,2\})$ denotes the transmit data sequence  sent from mobile user $i$ to the BS, $\bm{H}_{i}$ represents the channel fading gain between mobile user $i$ $(i\in\{1,2\})$ and the BS; and $\bm{W}$ is the additive white Gaussian noise (AWGN) process with zero mean and variance $\sigma^{2}$. 

We apply the hybrid SIC strategy~\cite{9151196} at the receiver to further improve the system performance.
In the traditional standard SIC approach, decoding and interference cancellation are done in a strict order. 
However, in hybrid SIC, some flexibility can be introduced and the SIC decoding order is opportunistically chosen.
In particular, by using the hybrid SIC strategy, the following n two possible SIC decoding orders should be considered:

\subsubsection{SIC Decoding Order 1} If mobile user 1 is decoded first, the signal-to-interference-plus-noise ratio (SINR), denoted by $\gamma_{1}^{(1)}$, is expressed as follows:
\begin{equation}\label{equation02}
\gamma_{1}^{(1)}=\frac{{\cal P}_{1}\left\|\bm{H}_{1}\right\|^{2}}{{\cal P}_{2}\left\|\bm{H}_{2}\right\|^{2}+\sigma^{2}}.
\end{equation}
After implementing SIC, the SINR for mobile user 2 is given by
\begin{equation}\label{equation02b}
	\gamma^{(1)}_{2}=\frac{{\cal P}_{2}\left\|\bm{H}_{2}\right\|^{2}}{\sigma^{2}}.
\end{equation}

\subsubsection{SIC Decoding Order 2} If mobile user 2 is decoded first, the SINR, denoted by $\gamma_{1}^{(2)}$, is expressed as follows:
\begin{equation}\label{equation02c}
	\gamma_{1}^{(2)}=\frac{{\cal P}_{1}\left\|\bm{H}_{1}\right\|^{2}}{{\cal P}_{1}\left\|\bm{H}_{1}\right\|^{2}+\sigma^{2}}.
\end{equation}
After implementing SIC, the SINR for mobile user 2 is given by
\begin{equation}\label{equation02b}
	\gamma^{(2)}_{2}=\frac{{\cal P}_{2}\left\|\bm{H}_{2}\right\|^{2}}{{\cal P}_{1}\left\|\bm{H}_{1}\right\|^{2}+\sigma^{2}}.
\end{equation}

\subsection{Channel Coding Rate}
Conventionally, Shannon's second theorem necessitates coding with infinite blocklength within the asymptotic regime. 
However, as highlighted earlier, in scenarios characterized by delay-sensitive QoS requirements, the application of Shannon's capacity formula becomes impractical. 
Considering the SIC decoding order $l$ $(l=1,2)$, we can derive the \textit{normal approximation} of the \textit{maximum achievable coding rate}, denoted by $R^{(l)}_{i}$, as follows~\cite{yury2010}:
\begin{align}\label{equation03}
R^{(l)}_{i}
\approx C(n,\gamma^{(l)}_{i})-\sqrt{\frac{V(n,\gamma^{(l)}_{i})}{n}}\left[\frac{Q^{-1}(\epsilon_{i})}{\log 2}\right],
\end{align}
where $n$ is the codeword length, $Q^{-1}(\cdot)$ denotes the inverse of $Q$-function, $\epsilon_{i}$ is the decoding error probability for mobile user $i$, and $C(n,{\cal P}_{i},\gamma^{(l)}_{i})$ and $V(n,\gamma^{(l)}_{i})$ are the channel capacity and channel dispersion, as expressed through the following derivations, respectively,~\cite{Polyanskiy2009}:
\begin{equation}
\begin{cases}
C(n,\gamma^{(l)}_{i})=\log_{2}\left(1+\gamma^{(l)}_{i}\right); \\
V(n,\gamma^{(l)}_{i})=1-\frac{1}{\left(1+\gamma^{(l)}_{i}\right)^{2}}.
\end{cases}
\end{equation}
%where $\gamma^{(l)}_{i}$ represents the signal-to-noise radio (SNR) between mobile user $i$ and the BS. Eq.~(\ref{equation03}) suggests that, for a blockcode with a finite length $n$ and SNR $\gamma^{(l)}_{i}$, the achievable data transmission rate is determined by the right-hand side of Eq.~(\ref{equation03}), ensuring that the decoding error probability remains no larger than $\epsilon_{i}$.
The utilization of Eq.~(\ref{equation03}) enables the analysis of finite-blocklength data transmission in NOMA system, taking into account both delay-bounded and error-rate bounded QoS constraints, which is detailed in the following section.

\section{QoS Guarantees with Statistical Delay and Error-Rate Bounds using $\epsilon$-Effective Capacity Through Using FBC} \label{sec:qos1}
%In this section, we first briefly review the effective capacity model and extend the expression of the effective capacity into a finite blocklength scenario. Accordingly, we define the concept of $\epsilon$-effective capacity under the dual constraints of statistical delay and error-rate bounded QoS.

\subsection{Preliminaries for $\epsilon$-Effective Capacity Under Statistical Delay/Error-Rate Bounded QoS}
Previous research has been dedicated to the examination of statistical delay-bounded QoS theory, as documented in~\cite{cheng2016}, to analyze queuing behavior in the context of statistical arrival/service processes. 
For our proposed uplink two-user FBC-enhanced NOMA model, we introduce varied delay-bounded QoS provisions for distinct links, establishing the basis for an innovative statistical delay-bounded QoS provisioning framework, which brings forth several new challenges.
Accordingly, by leveraging the large deviation principle (LDP) and subject to sufficient conditions, the queue length process $\widetilde{Q}_{i}(l)$ converges in distribution to a random variable $\widetilde{Q}_{i}(\infty)$, satisfying the following~\cite{cheng2016}:
\begin{equation}\label{equation40}
	-\lim_{\widetilde{Q}_{i,\text{th}}\rightarrow\infty}\frac{\log\left(\text{Pr}
		\left\{\widetilde{Q}_{i}(\infty)>\widetilde{Q}_{i,\text{th}}\right\}\right)}{\widetilde{Q}_{i,\text{th}}}=\theta_{i},
\end{equation}
where $\widetilde{Q}_{i,\text{th}}$ signifies the queuing overflow threshold and $\theta_{i}$ is defined as the \textit{delay-bounded QoS exponent}, indicating the exponential decay rate with respect to the delay-bounded QoS violation probabilities.
The delay-bounded QoS exponent serves as an indicator of the level of stringency with bounded delay as the threshold advances.
%Specifically, a sufficiently small delay-bounded QoS exponent enables the system to accommodate an arbitrarily prolonged delay, whereas a sufficiently large delay-bounded QoS exponent renders the system intolerant of any delay.
For a given designated delay threshold, the delay violation probability characterizes the tail behaviors of the queueing delay, which is a crucial factor in determining the fundamental performance-limits of statistical delay-bounded QoS.

Moreover, it is imperative to derive the power allocation policy, denoted as $\bm{\nu}^{(l)}_{i}\triangleq\bm{\nu}(n,\theta_{i},\gamma^{(l)}_{i})$, which is contingent upon the blocklength $n$, SINR $\gamma^{(l)}_{i}$, and the QoS exponent $\theta_{i}$. 
By employing the power allocation policy, the instantaneous transmit power of mobile user $i$ is expressed as ${\cal P}_{i}(\bm{\nu}^{(l)}) = \bm{\nu}^{(l)}_{i}\overline{{\cal P}}_{i}$, where $\overline{{\cal P}}_{i}$ signifies the mean transmit power for mobile user $i$.
Consequently, the following mean power constraint needs to be satisfied:
\begin{equation}\label{equation42}
	\int_{0}^{\infty}{\cal P}_{i}(\bm{\nu}^{(l)})\,p_{\Gamma}(\gamma^{(l)}_{i})d\gamma^{(l)}_{i}\leq\overline{{\cal P}}_{i}, \qquad \forall \,\, \theta_{i}>0
\end{equation}
where $p_{\Gamma}(\gamma^{(l)}_{i})$ is the probability density function (pdf) with respect to the SNR over Nakagami-$m$ fading channel, which is defined as
\begin{equation}\label{equation42a}
p_{\Gamma}(\gamma^{(l)}_{i})=\frac{(\gamma^{(l)}_{i})^{m-1}}{\Gamma(m)}\left(\frac{m}{\overline{\gamma}_{i}}\right)^{m}
\exp\left(-\frac{m\gamma^{(l)}_{i}}{\overline{\gamma}_{i}}\right),
\end{equation}
where $\overline{\gamma}_{i}$ denotes the average SNR between mobile user $i$ $(i\in\{1,2\})$ and the BS and $\Gamma(\cdot)$ represents the gamma function.

Given that the traditional effective capacity~\cite{cheng2016} only ensures statistical delay-bounded QoS constraints but does not account for reliability requirements arising from finite-blocklength data transmissions, this section introduces a concept in the finite blocklength regime, considering a target decoding error probability $\epsilon_{i}$ and a target delay-bounded QoS exponent $\theta_{i}$.

\textit{Definition 1:} For a target decoding error probability $\epsilon_{i}$,  the uplink $\epsilon$-effective capacity function, denoted by $EC^{(l,\epsilon)}(\theta_{i})$, is defined as the maximum constant arrival rate for our developed schemes in the finite blocklength regime that a given service process can sustain while adhering to the statistical delay and error-rate bounded QoS requirements. This is formally expressed as follows:
\begin{align}\label{equation017}
EC^{(l,\epsilon)}(\theta_{i})\triangleq-\frac{1}{n\theta_{i}}\log\left(\mathbb{E}_{\gamma^{(l)}_{i}}\left[\epsilon_{i}+\left(1-\epsilon_{i}\right)e^{- n\theta_{i}R^{(l)}_{i}}\right]\right),
\end{align}
where $R^{(l)}_{i}$ represents the maximum coding rate for mobile user $i$ $(i\in\{1,2\})$.
%Then, the uplink aggregate $\epsilon$-effective capacity, denoted by $EC^{\epsilon}(\theta)$, between the considered NOMA-MAC user pair and the BS in the finite blocklength regime as follows:
%\begin{align}\label{equation017a}
%EC^{\epsilon}(\theta)=-\sum\limits_{i=1}^{2}\frac{1}{\theta_{i}}\log\left(\mathbb{E}_{\gamma^{(l)}_{i}}\left[\left(\epsilon+\left(1-\epsilon\right)e^{- \theta_{i}R^{(l)}_{i}}\right)\right]\right).
%\end{align}

For our proposed uplink two-user NOMA system, we concentrate on maximizing the individual $\epsilon$-effective capacity given a target error probability $\epsilon_{i}$. 
Accordingly, using Eq.~(\ref{equation017}), the following optimization problem $\mathbf{P_{1}}$ aims to maximize the uplink $\epsilon$-effective capacity within the finite blocklength regime while upper-bounding both delay and error:
\begin{align}\label{equation020}
\mathbf{P_{1}}\!\!:\! \arg\max_{\{\bm{\nu}^{(l)},u_{l}\}}\!\left\{\!-\sum_{l=1}^{2}\frac{u_{l}}{n\theta_{i}}\!\log\!\left(\!\mathbb{E}_{\gamma^{(l)}_{i}}\!\!\left[\epsilon_{i}\!+\!\left(1\!-\!\epsilon_{i}\right)\!e^{- n\theta_{i}R^{(l)}_{i}}\!\right]\!\right)\!\!\right\}\!,
\end{align}
\begin{align}\label{equation019}
&\text{s.t.:} \, (a)\, \mathbb{E}_{\gamma^{(l)}_{i}}[{\cal P}_{i}(\bm{\nu}^{(l)})]\leq\overline{{\cal P}}_{i}; \nonumber \\
&\,(b)\!
%R^{(l)}_{i}\approx C(n,\gamma^{(l)}_{i})-\sqrt{\frac{V(n,\gamma^{(l)}_{i})}{n}}\left[\frac{Q^{-1}(\epsilon_{i})}{\log 2}\right]; \nonumber \\
\begin{cases}
\!R^{(1)}_{1}
\!\leq \min \left\{C\left(n,\gamma^{(1)}_{1}\right)-\sqrt{nV\left(n,\gamma^{(1)}_{1}\right)}Q^{-1}\left(\epsilon_{1}\right), \right.\\
\qquad\qquad\quad \left.{C\left(n,\gamma^{(1)}_{2}\right)-\sqrt{nV\left(n,\gamma^{(1)}_{2}\right)}Q^{-1}\left(\epsilon_{2}\right)}\right\};\\
\!R^{(1)}_{2}
\!\leq C\left(n,\gamma^{(1)}_{2}\right)\!-\!\sqrt{n V\left(n,\gamma^{(1)}_{2}\right)} Q^{-1}\!\left(\epsilon_{2}\right);\\
\!R^{(2)}_{1}
\!\leq C\left(n,\gamma^{(2)}_{1}\right)\!-\!\sqrt{n V\left(n,\gamma^{(2)}_{1}\right)} Q^{-1}\!\left(\epsilon_{1}\right);\\
\!R^{(2)}_{2}
\!\leq \min \left\{C\left(n,\gamma^{(2)}_{2}\right)-\sqrt{nV\left(n,\gamma^{(2)}_{2}\right)}Q^{-1}\left(\epsilon_{2}\right), \right.\\
\qquad\qquad\quad \left.{C\left(n,\gamma^{(2)}_{1}\right)-\sqrt{nV\left(n,\gamma^{(2)}_{1}\right)}Q^{-1}\left(\epsilon_{1}\right)}\right\};\\
\end{cases} \nonumber \\
&\,(c)\, n\geq n^{\text{th}};
\nonumber \\
&\,(d)\, \sum_{l=1}^{2}u_{l}=1,
\end{align}
where $n^{\text{th}}$ signifies the minimum blocklength required for constraint~$(c)$ specified by Eq.~(\ref{equation019}) to be satisfied and $u_{l}$ denotes a binary variable which controls the decoding order selection, i.e.,
\begin{equation}
	\begin{cases}
		u_{1}=1 \text{ and } u_{2}=0, \qquad  \text{if mobile user 1 is decoded first};\\
		u_{1}=0 \text{ and } u_{2}=1, \qquad  \text{if mobile user 2 is decoded first}.
	\end{cases}
\end{equation}

The problem  $\mathbf{P_{1}}$ is inherently non-convex and can be classified as a discrete optimization problem. To address this challenge effectively, the problem  $\mathbf{P_{1}}$ is decomposed into two sub-problems, each corresponding to a distinct SIC decoding order. 
These sub-problems are then solved concurrently at the BS. Based on the solutions obtained, the BS determines the optimal decoding order and communicates this decision to the receiver. Note that in practical communication systems, the BS possesses ample energy and computational resources, making the implementation of the proposed scheme feasible. The decomposition of problem  $\mathbf{P_{1}}$  is conducted as follows.

\subsection{Optimal Power Allocation Policy for Uplink FBC-Enhanced NOMA Under Statistical QoS Requirements}
Since the two questions sub-problems are remarkably
similar, they can be addressed by the similar approach.
Thus, we focus on analyzing scenario with the SIC decoding order 1.
\subsubsection{Optimal Power Allocation for Mobile User 2}
The non-convex maximization problem $\mathbf{P_{1}}$ given by Eq.~(\ref{equation020}) is converted into the following minimization problem $\mathbf{P_{2}}$:
\begin{align}\label{equation021}
&\mathbf{P_{2}}\!:
\!\arg\min_{\bm{\nu}^{(1)}_{2}}\!\vvast\{\!\mathbb{E}_{\gamma^{(1)}_{2}}\!\!\vvast[\epsilon_{2}\!+\!\left(1\!-\!\epsilon_{2}\right)\exp\!\vvast(\!\!-\!\!\vvast(\!\log_{2}\left(\!1\!+\!\bm{\nu}^{(1)}_{2}\gamma^{(1)}_{2}\!\right)\nonumber\\
&\,\,\times \theta_{2}-\sqrt{\frac{1}{n}\left(1-\frac{1}{(1+\bm{\nu}^{(1)}_{2}\gamma^{(1)}_{2})^{2}}\right)} 
 \left[\frac{Q^{-1}\left(\epsilon_{2}\right)}{\log 2}\right]\vvast)\!\vvast)\vvast]\vvast\}, 
%\nonumber\\
%=&\arg\min_{{\cal P}_{i}}\Bigg\{\mathbb{E}_{\gamma^{(l)}_{i}}\Bigg[\epsilon_{i}+\left(1-\epsilon_i{}\right)\frac{Q^{-1}(\epsilon_{i})}{\log 2}
%\nonumber\\
%\quad\times \Bigg( \left(1\!+\!\gamma^{(l)}_{i}\right)\!-\!\sqrt{\frac{1}{n}\!\left(1\!-\!\frac{1}{(1\!+\!\gamma^{(l)}_{i})^{2}}\right)} \Bigg)^{-\beta_{i}}\Bigg]\Bigg\},
\end{align}
while satisfying the same constraints specified in Eq.~(\ref{equation019}). 
Then, to address the optimization problem above, we first explore the monotonicity of the uplink $\epsilon$-effective capacity function concerning the blocklength, as outlined in the following lemma.

\begin{lemma} \label{lemma02}
	For a target decoding error probability $\epsilon_{2}\in(0,1/2)$ and the SNR $\gamma^{(1)}_{2}$, the uplink $\epsilon$-effective capacity between the link from mobile user 2 to the BS is monotonically increasing with respect to the blocklength $n$.
\end{lemma}
\begin{IEEEproof}
To establish the monotonicity considering the uplink $\epsilon$-effective capacity with respect to the blocklength $(n>0)$, we initially introduce the following function: 
\begin{align}\label{equation014}
&F(n, \epsilon_{2},\gamma^{(1)}_{2})\triangleq \,n\log_{2}\left(1+\bm{\nu}^{(1)}_{2}\gamma^{(1)}_{2}\right)
\nonumber\\
&\quad-\!\sqrt{n\left(1-\frac{1}{(1+\bm{\nu}^{(1)}_{2}\gamma^{(1)}_{2})^{2}}\right)}   \left[\frac{Q^{-1}\left(\epsilon_{2}\right)}{\log 2}\right].
\end{align}
Subsequently, the first-order derivative of the auxiliary function $F(n,\epsilon_{2},\gamma^{(1)}_{2})$ is expressed as follows:
\begin{align}
&\frac{\partial F(n,\epsilon_{2},\gamma^{(1)}_{2})}{\partial n}\nonumber\\
&\quad=\log_{2}\!\left(1\!+\!\bm{\nu}^{(1)}_{2}\gamma^{(1)}_{2}\right)\!-\!\sqrt{1\!-\!\frac{1}{(1\!+\!\bm{\nu}^{(1)}_{2}\gamma^{(1)}_{2})^{2}}}\left[\! \frac{Q^{-1}\left(\epsilon_{2}\right)}{2\sqrt{n}\log 2}\right] \nonumber\\
&\quad=\frac{1}{2}\left(\log_{2}\left(1+\bm{\nu}^{(1)}_{2}\gamma^{(1)}_{2}\right)+\frac{R^{(1)}_{2}}{n}\right),
\end{align}
which is due to the following:
\begin{align}
&\sqrt{\frac{1}{n}\left(1\!-\!\frac{1}{(1+\bm{\nu}^{(1)}_{2}\gamma^{(1)}_{2})^{2}}\right)}\left[\frac{Q^{-1}\left(\epsilon_{2}\right)}{\log 2}\right]
\nonumber\\
&\qquad=\log_{2}\!\left(\!1\!+\!\bm{\nu}^{(1)}_{2}\gamma^{(1)}_{2}\right)\!-\!\frac{R^{(1)}_{2}}{n}.
\end{align}
As a result, we can establish that $\frac{\partial F(n,\epsilon_{2},\gamma^{(1)}_{2})}{\partial n}>0$. 
Consequently, $F(n,\epsilon_{2},\gamma^{(1)}_{2})$ monotonically increases with respect to $n$ when $n>0$.
Subsequently, the correlation between the uplink $\epsilon$-effective capacity and $F(n,\epsilon_{2},\gamma^{(1)}_{2})$ is delineated as follows:
\begin{equation}\label{equation022}
EC^{(1,\epsilon)}(\theta_{2})\!=\!-\frac{1}{\theta_{2}}\log\mathbb{E}_{\gamma^{(1)}_{2}}\!\left[\epsilon_{2}\!+\!\left(1\!-\!\epsilon_{2}\right)e^{-\theta_{2}\frac{F(n,\epsilon_{2},\gamma^{(1)}_{2})}{n}}\right].
\end{equation}
Hence, it is necessary to examine the monotonicity of $\frac{F(n,\epsilon_{2},\gamma^{(1)}_{2})}{n}$, and it is expressed as follows:
\begin{align}\label{equation220}
\frac{\partial\frac{F(n,\epsilon_{2},\gamma^{(1)}_{2})}{n}}{\partial n}=&\frac{nF'_{n}(n,\epsilon_{2},\gamma^{(1)}_{2})-F(n,\epsilon_{2},\gamma^{(1)}_{2})}{n^{2}} \nonumber \\
%=&\frac{n\log_{2}\left(1+\gamma^{(l)}_{i}\right)-\sqrt{1-\frac{1}{(1+\gamma^{(l)}_{i})^{2}}}\frac{\sqrt{n}Q^{-1}(\epsilon)}{2\log 2}}{n^{2}} \nonumber \\
%\!\!\!\!\!\!\!\!\!\!\!\!\!\!-\!\frac{n\log_{2}\left(1\!+\!\gamma^{(l)}_{i}\right)\!-\!\sqrt{n\!\left(\!1\!-\!\frac{1}{(1+\gamma^{(l)}_{i})^{2}}\right)}\frac{Q^{-1}(\epsilon)}{\log 2}}{n^{2}} \nonumber\\
=&\frac{\sqrt{n}Q^{-1}\left(\epsilon_{2}\right)\sqrt{1-\frac{1}{(1+\bm{\nu}^{(1)}_{2}\gamma^{(1)}_{2})^{2}}}}{2n^{2}\log 2}>0,
\end{align}
where $F'_{n}(n,\epsilon_{2},\gamma^{(1)}_{2})\triangleq{\partial F(n,\epsilon_{2},\gamma^{(1)}_{2})}/{\partial n}$ while $0<\epsilon_{2}<1/2$.
As the exponential function $e^{-x}$ monotonically decreases while $\log(x)$ monotonically increases in terms of $x$, it can be observed that the uplink $\epsilon$-effective capacity function $EC^{(1,\epsilon)}(\theta_{2})$ monotonically increases with respect to $F(n,\epsilon_{2},\gamma_{2})$. 
Consequently, it implies that the uplink $\epsilon$-effective capacity monotonically increases with respect to $n$. This completes the proof of Lemma~\ref{lemma02}.
\end{IEEEproof}

\indent\textit{Remarks on Lemma~\ref{lemma02}:} Lemma~\ref{lemma02} delineates the monotonic behavior of the uplink $\epsilon$-effective capacity with respect to the $n$ through employing FBC. 
Consequently, the results form the basis for addressing the optimization problem $\mathbf{P_{2}}$ and assessing the efficacy of our proposed methodologies aimed at achieving QoS provisioning within the finite blocklength regime, as expounded in the following theorem.

\begin{theorem}\label{theorem02}
For the decoding error probability $\epsilon_{2}\in(0,1/2)$, the optimization problem $\mathbf{P_{2}}$ as defined in Eq.~(\ref{equation021}) is strictly convex within the space spanned by $(n, \bm{\nu}^{(1)}_{2})$ when $n> \max \left\{\left(n^{\text{rt}}\right)^{2}, n^{\text{th}}\right\}$. Here, $n^{\text{th}}$ is determined by the constraint~(c) presented by Eq.~(\ref{equation019}), and
\begin{align}\label{equation023}
\begin{cases}
\!n^{\text{rt}}&\!\!\!\!\!\triangleq\!\left[\!\frac{Q^{-1}\left(\epsilon_{2}\right)}{8}\right]\!\!\left(\frac{4}{\overline{a}\overline{b}}\!-\!\frac{\overline{b}}{\overline{a}}\right)\!+\!\sqrt{\frac{1}{16}\!\left(\frac{4}{\overline{a}\overline{b}}\!-\!\frac{\overline{b}}{\overline{a}}
\!\right)^{\!2}\!+\!3\left[Q^{-1}\!\left(\epsilon_{2}\right)\!\right]^{2}}\\
&\times \left[\frac{Q^{-1}\left(\epsilon_{2}\right)}{2}\right]; \\
\!\overline{a}&\!\!\!\!\!\triangleq 1+\bm{\nu}^{(1)}_{2}\gamma^{(1)}_{2}; \\
\!\overline{b}&\!\!\!\!\!\triangleq\sqrt{\bm{\nu}^{(1)}_{2}\gamma^{(1)}_{2}\left(\bm{\nu}^{(1)}_{2}\gamma^{(1)}_{2}+2\right)}.
\end{cases}
\end{align}
\end{theorem}
\begin{IEEEproof}
	The proof of Theorem~\ref{theorem02} is in Appendix A.
\end{IEEEproof}

\begin{figure}[!t]
	\centering
	\includegraphics[scale=0.53]{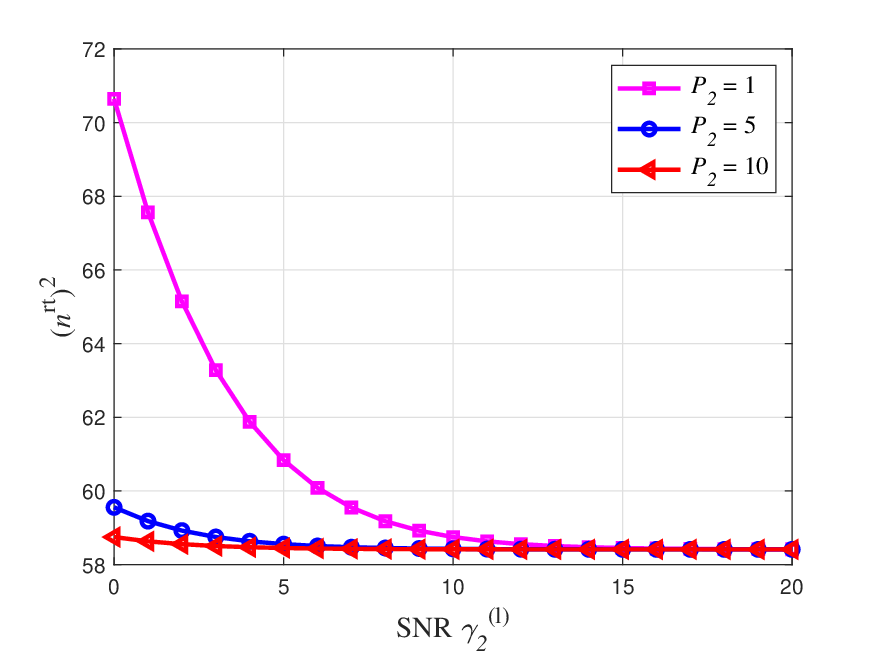}
	\caption{The relationship between $(n^{\text{rt}})^{2}$ and the SNR $\gamma^{(1)}_{2}$  for our proposed  uplink two-user NOMA based scheme under FBC.}
	\label{fig:04}
\end{figure}

\indent\textit{Remarks on Theorem~\ref{theorem02}:} Theorem~\ref{theorem02} shows that the optimization problem $\mathbf{P_{2}}$ given in Eq.~(\ref{equation021}) is convex within the space defined by $(n,\bm{\nu}^{(l)})$ under the conditions $0<\epsilon_{2}<1/2$ and $n> \max \left\{\left(n^{\text{rt}}\right)^{2}, n^{\text{th}}\right\}$. Accordingly, there is a unique solution when $n> \left\{\left(n^{\text{rt}}\right)^{2}, n^{\text{th}}\right\}$ and $\epsilon_{2}\in(0,1/2)$. 
To help demonstrate the dynamics structures and connections relationship between $(n^{\text{rt}})^{2}$ and the SNR $\gamma^{(1)}_{2}$, by fixing the decoding error probability $\epsilon_{2}$ at $10^{-6}$, Fig.~\ref{fig:04} depicts $(n^{\text{rt}})^{2}$ with respect to the SNR $\gamma^{(1)}_{2}$ for the proposed uplink two-user FBC-enhanced NOMA schemes. 
Fig.~\ref{fig:04} demonstrates that the value of $(n^{\text{rt}})^{2}$ diminishes with an increase in transmit power ${\cal P}_{2}$.
Setting the transmit power ${\cal P}_{2}=1$, it is evident that the value of $(n^{\text{rt}})^{2}$ decreases from 70.64 to 58.42 when $\gamma^{(1)}_{2}$ varies from 0 to 20 dB. 
Given that the authors of~\cite{yury2010} have demonstrated the accuracy of data transmission rates even when $n^{\text{th}}$ is as short as 100, which is greater than $(n^{\text{rt}})^{2}$, the constraint in Eq.~(\ref{equation052}) is automatically satisfied for $n>n^{\text{th}}$.
%Moreover, the lower bound of the blocklength $\left(n^{\text{rt}}\right)^{2}$ depends on the error probability $\epsilon$, SNR $\gamma^{(l)}_{i}$, and power ${\cal P}_{i}$. This implies that the upper bound of the blocklength can be large enough for reliable data transmissions $0<\epsilon<10^{-3}$ (i.e., reliability requirement of traditional  services).

Correspondingly, the optimal power allocation policy for our developed FBC-enhanced NOMA system can be obtained through the following theorem.	

\begin{theorem} \label{theorem03}
Under the conditions $0<\epsilon_{2}<1/2$ and $n> \max \left\{\left(n^{\text{rt}}\right)^{2}, n^{\text{th}}\right\}$ while considering the high SNR region, the optimal power allocation policy, denoted by $\bm{\nu}^{(1,\text{opt})}_{2}$, for maximizing the $\epsilon$-effective capacity across the uplink data transmissions under the constraints in terms of the statistical delay and error-rate bounded QoS over the developed FBC-enhanced NOMA system is given as follows:
\begin{equation}\label{equation035a}
\bm{\nu}^{(1,\text{opt})}_{2}\!=\!
\begin{cases}
\!\frac{\left(\beta_{2}\right)^{\frac{1}{\beta_{2}+1}}\!\!\left[(\!1-\epsilon_{2})e^{\frac{\theta_{2}Q^{-1}\left(\epsilon_{2}\right)}{\sqrt{n}\log 2}}\!\right]^{\!\frac{1}{\beta_{2}+1}}}{\left(\gamma^{(1)}_{2}\right)^{\frac{\beta_{2}}{\beta_{2}+1}}\left(\widetilde{\lambda}^{\text{opt}}_{2}\right)^{\frac{1}{\beta_{2}+1}}}, 
\qquad  \gamma^{(1)}_{2}\geq \frac{\widetilde{\lambda}^{\text{opt}}_{2}}{\beta_{2}};  \\
\!0, \qquad\qquad\qquad\qquad\qquad\qquad\qquad \gamma^{(1)}_{2}<\frac{\widetilde{\lambda}^{\text{opt}}_{2}}{\beta_{2}},
\end{cases}
\end{equation}
where $\beta_{2}\triangleq\theta_{2}/\log2$ and $\widetilde{\lambda}^{\text{opt}}_{2}$ denotes the optimal Lagrange multiplier, which is numerically determined through plugging Eq.~(\ref{equation035a}) in constraint (a) as given by Eq.~(\ref{equation019}).
\end{theorem}

\begin{IEEEproof}
		The proof of Theorem~\ref{theorem03} is in Appendix B.
\end{IEEEproof}

%\indent\textit{Remarks on Theorem~\ref{theorem03}:} The results established in Theorem~\ref{theorem03} provide valuable engineering guidance for the practical design, analysis, and evaluation of our proposed performance modeling formulations, particularly in the context of statistical delay and error-rate bounded QoS provisioning.

\subsubsection{Optimal Power Allocation for Mobile User 1}
According to the maximum achievable coding rate for mobile user 1, which is specified in constraint (b) in Eq.~\eqref{equation019}, i.e., 
\begin{align}
R^{(1)}_{1}
\!\leq \min \left\{C\left(n,\gamma^{(1)}_{1}\right)-\sqrt{nV\left(n,\gamma^{(1)}_{1}\right)}Q^{-1}\left(\epsilon_{1}\right), \right. \nonumber\\
\qquad\qquad\quad \left.{C\left(n,\gamma^{(1)}_{2}\right)-\sqrt{nV\left(n,\gamma^{(1)}_{2}\right)}Q^{-1}\left(\epsilon_{2}\right)}\right\}
\end{align}
we can maximize the $\epsilon$-effective capacity for mobile user 1 while considering both the perfect and imperfect SIC cases listed as follows:

\underline{\textbf{Case 1}}: With perfect SIC,  the optimization problem that maximizes the $\epsilon$-effective capacity for mobile user 1 is formulated as follows:
\begin{align}\label{equation021a}
\mathbf{P_{3}}:&
\arg\min_{\bm{\nu}^{(1)}_{1}}\!\Bigg\{\!\mathbb{E}_{\gamma^{(1)}_{1}}\!\Bigg[\!\epsilon_{1}\!+\!\left(1\!-\!\epsilon_{1}\right)\exp\!\Bigg(\!\!-\theta_{1}\Bigg( \log_{2}\!\left(\!1\!+\!\bm{\nu}^{(1)}_{1}{\gamma}^{(1)}_{1}\!\right) \nonumber\\
&\qquad\qquad\,\,-\!\sqrt{\frac{1}{n}\!\left(\!1\!-\!\frac{1}{(1\!+\!\bm{\nu}^{(1)}_{1}{\gamma}^{(1)}_{1})^{2}}\right)}\!\! \left[\!\frac{Q^{-1}\left(\epsilon_{1}\right)}{\log 2}\!\right]\!\Bigg)\!\Bigg)\Bigg]\Bigg\}, 
\end{align}
while satisfying the same constraints specified by Eq.~(\ref{equation019}). 
Similarly, as stated when deriving the optimal power allocation policy for mobile user 2, we can prove that $\mathbf{P_{3}}$ given in Eq.~(\ref{equation021a}) is strictly convex on the space defined by $(n, \bm{\nu}_{1})$ given the similar constraints in Theorem~\ref{theorem02}.
Correspondingly, the optimal power allocation policy, denoted by $\bm{\nu}_{1}^{(1,\text{opt})}$, for mobile user 1 is obtained as in the following equation:
\begin{equation}\label{equation0035}
\bm{\nu}_{1}^{(1,\text{opt})}\!=\!
\begin{cases}
\!\frac{\left(\beta_{1}\right)^{\frac{1}{\beta_{1}+1}}\!\left[\!(1-\epsilon_{1})e^{\frac{\theta_{1}Q^{-1}\left(\epsilon_{2}\right)}{\sqrt{n}\log 2}}\!\right]^{\frac{1}{\beta_{1}+1}}}{\left({\gamma}^{(1)}_{1}\right)^{\frac{\beta_{1}}{\beta_{1}+1}}\left(\widetilde{\lambda}^{\text{opt}}_{1}\right)^{\frac{1}{\beta_{1}+1}}}, 
\quad\,\,\,  {\gamma}^{(1)}_{1}\geq \frac{\widetilde{\lambda}^{\text{opt}}_{1}}{\beta_{1}};  \\
\!0, \qquad\qquad\qquad\qquad\qquad\qquad\qquad\! {\gamma}^{(1)}_{1}<\frac{\widetilde{\lambda}^{\text{opt}}_{1}}{\beta_{1}},
\end{cases}
\end{equation}
where $\beta_{1}\triangleq\theta_{1}/\log2$ and $\widetilde{\lambda}^{\text{opt}}_{1}$ denotes the optimal Lagrange multiplier, which is numerically determined through plugging Eq.~(\ref{equation035a}) in constraint (a) as given by Eq.~(\ref{equation019}).

\underline{\textbf{Case 2}}: In the presence of imperfect SIC, we formulate the problem aiming to maximize the $\epsilon$-effective capacity for mobile user 1, which is given as follows:
\begin{align}\label{equation021b}
\mathbf{P_{4}}:&
\arg\min_{\bm{\nu}^{(1)}_{2}}\!\Bigg\{\mathbb{E}_{\gamma^{(1)}_{2}}\!\Bigg[\!\epsilon_{1}\!+\!\left(1\!-\!\epsilon_{1}\right)\exp\!\Bigg(\!\!-\theta_{1}\!\Bigg(\!\! \log_{2}\!\left(1\!+\!\bm{\nu}^{(1)}_{2}{\gamma}^{(1)}_{2}\!\right) \nonumber\\
&\qquad\qquad\!-\!\sqrt{\frac{1}{n}\left(1\!-\!\frac{1}{(1\!+\!\bm{\nu}^{(1)}_{2}{\gamma}^{(1)}_{2})^{2}}\right)}\!\! \left[\frac{Q^{-1}\left(\epsilon_{2}\right)}{\log 2}\!\right]\!\Bigg)\!\Bigg)\Bigg]\!\Bigg\}, 
\end{align}
subject to the same constraints specified by  Eq.~(\ref{equation019}). 
Likewise,  the optimal power allocation policy can be obtained for solving $\mathbf{P_{4}}$ defined by Eq.~(\ref{equation021b}).

\subsection{Maximum Uplink $\epsilon$-Effective Capacity in FBC-Enhanced NOMA}

Based on the developed optimal power allocation policy in the previous section, the closed-form expression of the maximum uplink $\epsilon$-effective capacity $EC^{(l,\epsilon)}_{\text{max}}(\theta_{i})$ for mobile user $i$ in the high SNR region, as outlined in the theorem that follows.

\begin{theorem} \label{theorem04}
Considering the constraints $n> \max \left\{\left(n^{\text{rt}}\right)^{2}, n^{\text{th}}\right\}$ and $\epsilon_{i}\in(0,1/2)$, the maximum uplink $\epsilon$-effective capacity $EC^{(l,\epsilon)}_{\text{max}}(\theta_{i})$ $(i\in\{1,2\})$ for FBC-enhanced NOMA system is expressed as follows:
	\begin{align}\label{equation036}
	EC^{(l,\epsilon)}_{\text{max}}(\theta_{i})\!=&\!-\frac{1}{\theta_{i}}\!\log\vast\{\!\mathbb{E}_{\gamma^{(l)}_{i}}\vast[\epsilon_{i}+\left(1\!-\!\epsilon_{i}\right)\exp\!\vast(
	\!\left[\frac{\theta_{i}Q^{-1}(\epsilon_{i})}{\log 2}\right]\nonumber \\
	&\times\!\! \sqrt{\!\frac{1}{n}\!\left(\!1\!-\!\frac{1}{\left[\bm{\nu}^{(l,\text{opt})}_{i}\gamma^{(l)}_{i}\right]^{2}}\!\right)}\vast)\!
	\!\left(\bm{\nu}^{(l,\text{opt})}_{i}\gamma^{(l)}_{i}\right)^{-\beta_{i}}\!\vast]\!\vast\} \nonumber \\
=&-\frac{1}{\theta_{i}}\log\Bigg\{\mathbb{E}_{\gamma^{(l)}_{i}}\Bigg[\epsilon_{i}+\left(1\!-\!\epsilon_{i}\right)\exp\!\Bigg(
\!\left[\frac{\theta_{i}Q^{-1}(\epsilon_{i})}{\sqrt{n}\log 2}\right]\nonumber \\
&\times \sqrt{1-\left(\widetilde{a}\beta_{i}\gamma^{(l)}_{i}\right)^{-\frac{2}{\beta_{i}+1}}}\Bigg)
\!\left(\widetilde{a}\beta_{i}\gamma^{(l)}_{i}\right)^{-\frac{\beta_{i}}{\beta_{i}+1}}\!\Bigg]\!\Bigg\}, 
%	=&\!-\!\frac{1}{\theta_{i}}\!\log\left(1\!-\!\epsilon\right)\!-\!\frac{1}{\theta_{i}}\!\log\!\Bigg\{\mathbb{E}_{\gamma^{(l)}_{i}}\Bigg[\!\exp\!\Bigg\{
%	\frac{\theta_{i}Q^{-1}(\epsilon)}{\log 2}\nonumber\\
%	&\times \sqrt{\frac{1}{n}\left(1-\left(\widetilde{a}\gamma^{(l)}_{i}\right)^{-\frac{2}{\beta_{i}+1}}\right)}\Bigg\}\!
%	\left(\widetilde{a}\gamma^{(l)}_{i}\right)^{-\frac{\beta_{i}}{\beta_{i}+1}}\!\Bigg]\!\Bigg\},
	\end{align}
	and the closed-form expression in terms of the maximum uplink $\epsilon$-effective capacity is approximately determined as in the following equation:
	\begin{align}\label{equation036a}
	EC^{(l,\epsilon)}_{\text{max}}(\theta_{i})&\approx
\!-\!\frac{1}{\theta_{i}}\Bigg\{\!\!\log\!\left\{\!\!\left(1\!-\!\epsilon_{i}\right)\!
\left(\widetilde{a}\beta_{i}\overline{\gamma}_{i}\right)^{-\frac{\beta_{i}}{\beta_{i}+1}}\!\Gamma\!\left(\!\frac{1}{\beta_{i}\!+\!1},\!\frac{\widetilde{\lambda}^{\text{opt}}_{i}}{\beta_{i}}\right)\!\!\right\}
\nonumber\\
&+\!\left[\frac{\theta_{i}Q^{-1}(\epsilon_{i})}{\sqrt{n}\log 2}\right]\!
\Bigg[\!1\!-\!\frac{\left(\widetilde{a}\beta_{i}\overline{\gamma}_{i}\right)^{-\frac{2}{\beta_{i}+1}}}{2} \!\Gamma\left(\!\frac{\beta_{i}-1}{\beta_{i}\!+\!1},\!\frac{\widetilde{\lambda}^{\text{opt}}_{i}}{\beta_{i}}\!\right)
\nonumber\\
&\times - \frac{\left(\widetilde{a}\beta_{i}\overline{\gamma}_{i}\right)^{-\frac{4}{\beta_{i}+1}}}{8} 
\Gamma\left(\frac{\beta_{i}-3}{\beta_{i}+1},\frac{\widetilde{\lambda}^{\text{opt}}_{i}}{\beta_{i}}\right)
\nonumber\\
&-\sum_{\ell=3}^{\infty}\frac{(2\ell-3)!}{\ell!(\ell-2)!2^{2\ell-2}} \left(\widetilde{a}\beta_{i}\overline{\gamma}_{i}\right)^{-\frac{2\ell}{\beta_{i}+1}}  \nonumber\\
&\times
\Gamma\left(\frac{\beta_{i}-2\ell+1}{\beta_{i}+1},\frac{\widetilde{\lambda}^{\text{opt}}_{i}}{\beta_{i}}\right)\Bigg]\Bigg\},
	\end{align}
	where $\Gamma(\cdot,\cdot)$ denotes the upper incomplete gamma function and 
	\begin{align}\label{equation038}
	\widetilde{a}\triangleq\frac{(1-\epsilon_{i})\exp\left(\frac{\theta_{i}Q^{-1}(\epsilon_{i})}{\sqrt{n}\log 2}\right)}{\lambda^{\text{opt}}_{i}}.
	\end{align}
\end{theorem}

\begin{IEEEproof}	
	The proof of Theorem~\ref{theorem04} is in Appendix C.
\end{IEEEproof}

\indent\textit{Remarks on Theorem~\ref{theorem04}:} Theorem~\ref{theorem04} defines, creates, and analyzes the maximum uplink $\epsilon$-effective capacity $EC^{(l,\epsilon)}_{\text{max}}(\theta_{i})$ using our developed optimal power allocation policies in the high SNR region. 
The closed-form expressions of the maximum $\epsilon$-effective capacity given in Theorem~\ref{theorem04} can be applied to define the $\epsilon$-effective energy efficiency function in the next section.
%In addition, the maximum $\epsilon$-effective capacity function can be employed to characterize the asymptotic behaviors and performance metrics concerning the QoS exponent and the blocklength $n$, as detailed in Section~\ref{sec:qos2}.

\section{Statistical QoS Guarantees Through $\epsilon$-Effective Energy Efficiency With Unknown CSI in the Finite Blocklength Regime} \label{sec:qos2}
In scenarios where the CSI is unavailable at the transmitter, the implementation of a dynamic power allocation policy becomes infeasible due to the absence of instantaneous CSI. 
In this context, the consideration shifts towards an average power allocation policy. Consequently, we formulate the total power consumption for the data transmission between the uplink mobile user $i$ and the BS, denoted by ${\cal P}_{i,\text{o}}$, as follows~\cite{cheng2016}:
\begin{equation}
{\cal P}_{i,\text{o}}=\eta_{i}\overline{{\cal P}}_{i}+{\cal P}_{c},
\end{equation}
where $\eta_{i}$ signifies average transmit power efficiency, which increases proportionally with the average transmit power and ${\cal P}_{c}$ refers to the circuit power consumption at the mobile user, a parameter unaffected by average transmit power $\overline{{\cal P}}_{i}$.
Then, based on the definition of the $\epsilon$-effective  capacity in the previous section, we can the define the concept concerning the $\epsilon$-effective energy efficiency as follows. 

\textit{Definition 2:} For the target decoding error probability $\epsilon_{i}$, the uplink $\epsilon$-effective energy efficiency, denoted by  $EE^{(l,\epsilon)}(\theta_{i})$, is defined as the totally delivered bits per unit energy considering the constraints in terms of the statistical delay and error-rate, which is formally expressed as follows:
\begin{align}\label{equation18}
EE^{(l,\epsilon)}(\theta_{i})&\triangleq\frac{EC^{(l,\epsilon)}(\theta_{i})}{{\cal P}_{i,\text{o}}}.
%\nonumber\\
%&=-\frac{\log\left(\mathbb{E}_{\gamma^{(l)}_{i}}\left[\epsilon_{i}+\left(1-\epsilon_{i}\right)e^{- \theta_{i}R^{(l)}_{i}}\right]\right)}{\theta_{i}{\cal P}_{i,\text{o}}}.
%=\!-\!\frac{1}{\theta_{i}\!\left(\eta_{i}\overline{{\cal P}}_{i}+{\cal P}_{c}\right)}\log\!\Bigg(\mathbb{E}_{\gamma^{(l)}_{i}}\!\Bigg[\!\left(1\!-\!\epsilon\right)\exp\!\Bigg\{\!\!-\!\theta_{i}\nonumber\\
%\quad\times\! \Bigg(\! \log_{2}\left(1\!+\!\gamma^{(l)}_{i}\right)\!-\sqrt{\frac{1}{n}\!\left(\!1\!-\!\frac{1}{(1+\gamma^{(l)}_{i})^{2}}\!\right)}\nonumber\\
%\quad\times\frac{Q^{-1}(\epsilon)}{\log 2}  \Bigg)\!\Bigg\}\!\Bigg]\Bigg).
\end{align}

\subsection{Uplink $\epsilon$-Effective Energy Efficiency Maximization With Unknown CSI Through Applying FBC}
We apply the average power allocation policy, denoted by $\overline{\bm{\nu}}^{(l)}=[\bm{\nu}^{(l)}_{1},\bm{\nu}^{(l)}_{2}]$.
Accordingly, the average transmit power of mobile user $i$ becomes $\overline{{\cal P}}_{i}(\overline{\bm{\nu}}^{(l)}) =\overline{\bm{\nu}}^{(l)}_{i}\overline{{\cal P}}_{i}$.
As a result, in cases where the transmitter possesses only the knowledge of the CSI distribution, commonly referred to as channel distribution information (CDI)~\cite{10066170}, the utilization of Eqs.~(\ref{equation03}) and~(\ref{equation18}) enables the formulation of the optimization problem $\mathbf{P_{5}}$. 
This problem is aimed at maximizing the uplink $\epsilon$-effective energy efficiency, denoted by $EE^{(l,\epsilon)}_{\text{max}}(\theta_{i})$, under the constraints with respect to statistical delay and error-rate QoS provisioning in the finite blocklength regime, as expressed below:
\begin{align}\label{equation20}
\mathbf{P_{5}}: &EE^{(l,\epsilon)}_{\text{max}}(\theta_{i})\nonumber\\
&\quad\!\! =\!\arg\max_{\overline{\bm{\nu}}^{(l)}_{i}}\left\{\!-\frac{\log\!\left(\mathbb{E}_{\gamma^{(l)}_{i}}\!\left[\epsilon_{i}\!+\!\left(1\!-\!\epsilon_{i}\right)e^{- \theta_{i}R^{(l)}_{i}}\right]\right)}{\theta_{i}{\cal P}_{i,\text{o}}}\right\},
\end{align}
\begin{align}\label{equation19}
&\text{s.t. :} \,\, (a)\,\,  0<\overline{{\cal P}}_{i}(\overline{\bm{\nu}}^{(l)})\leq\overline{{\cal P}}_{i}^{\text{th}}; \nonumber\\
&\,(b)\,\, R^{(l)}_{i}\leq\log_{2}\!\left(1\!+\!\overline{\bm{\nu}}^{(l)}_{i}\gamma^{(l)}_{i}\right)-\sqrt{\frac{1}{n}\!\left(\!1\!-\!\frac{1}{(1\!+\!\overline{\bm{\nu}}^{(l)}_{i}\gamma^{(l)}_{i})^{2}}\!\right)}   \nonumber \\
&\qquad\qquad\quad \times Q^{-1}(\epsilon_{i})\log_{2}(e);\nonumber \\
&\,(c)\,\, n\geq n^{\text{th}},
\end{align}
where $\overline{{\cal P}}_{i}^{\text{th}}$ denotes the maximum allowed average transmit power. 
To analyze  the maximization problem  $\mathbf{P_{5}}$ given by Eq.~(\ref{equation20}), first, we need to analyze the monotonicity of the $\epsilon$-effective energy efficiency as detailed in the following lemma.

\begin{lemma}  \label{lemma05}
	The maximization problem $\mathbf{P_{5}}$ exhibits strict quasi-concavity in terms of $\overline{\bm{\nu}}^{(l)}_{i}$ under the constraints $n> \max \left\{\left(n^{\text{rt}}\right)^{2}, n^{\text{th}}\right\}$ and $\epsilon_{i}\in(0,1/2)$. 
\end{lemma}
\begin{IEEEproof}
	First, we define the super-level set ${\cal S}_{\epsilon}$ concerning $EE^{(l,\epsilon)}_{\text{max}}(\theta_{i})$ as follows:
	\begin{equation}\label{equation16}
	{\cal S}_{\alpha}=\left\{\overline{\bm{\nu}}^{(l)}_{i}\geq0|EE^{(l,\epsilon)}_{\text{max}}(\theta_{i})\geq\alpha
	\right\},
	\end{equation}
	for $\alpha\in\mathbb{R}$. The authors in~\cite{SB2004} demonstrated that $EE^{(l,\epsilon)}_{\text{max}}(\theta_{i})$ is quasiconcave in $\overline{\bm{\nu}}^{(l)}_{i}$ if ${\cal S}_{\alpha}$ is convex with respect to $\alpha\in\mathbb{R}$. 
	If $\alpha<0$, there exists no solution for $EE^{(l,\epsilon)}_{\text{max}}(\theta_{i})=\alpha$. 
	On the other hand, if $\alpha\geq0$, the superset ${\cal S}_{\alpha}$ can be rewritten as follows:
	\begin{equation}
	{\cal S}_{\alpha}=\left\{\overline{\bm{\nu}}^{(l)}_{i}\geq0|\alpha\left(\eta_{i}\overline{{\cal P}}_{i}(\overline{\bm{\nu}}^{(l)})+{\cal P}_{c}\right)-EC^{(l,\epsilon)}_{\text{max}}(\theta_{i})\leq 0
	\right\}.
	\end{equation}
	Based on the results in Theorem~\ref{theorem02}, we can then establish that ${\cal S}_{\alpha}$ is strictly convex with respect to $\overline{\bm{\nu}}^{(l)}_{i}$. 
	Therefore, the maximization problem $\mathbf{P_{5}}$ given by Eq.~(\ref{equation20}), being the ratio of a concave function over a non-negative affine function in $\overline{\bm{\nu}}^{(l)}_{i}$, exhibits strict quasi-concavity in terms of $\overline{\bm{\nu}}^{(l)}_{i}$, concluding the proof of Lemma~\ref{lemma05}.
\end{IEEEproof}

\indent\textit{Remarks on Lemma~\ref{lemma05}:} Lemma~\ref{lemma05} indicates that for a strictly quasiconcave function, there is always a unique global maximum for $\mathbf{P_{5}}$ when $n> \max \left\{\left(n^{\text{rt}}\right)^{2}, n^{\text{th}}\right\}$ and $\epsilon_{i}\in(0,1/2)$. 
Thus, Lemma~\ref{lemma05} ensures both the existence and uniqueness of the global maximum and unveils the differentiability of the $\epsilon$-effective energy efficiency function.
To prove that the maximum is always achieved at a finite average transmit power, by substituting constraint (b) from  Eq.~(\ref{equation19}) into Eq.~(\ref{equation20}),  we solely focus on addressing the average power constraint (a) outlined in Eq.~(\ref{equation19}) to solve $\mathbf{P_{5}}$, as detailed in the following theorem.

\begin{theorem}  \label{theorem08}
	When $n> \max \left\{\left(n^{\text{rt}}\right)^{2}, n^{\text{th}}\right\}$ and $\epsilon_{i}\in(0,1/2)$, the following claims hold true.
	
	\underline{Claim 1:} The uplink $\epsilon$-effective capacity $EC^{(l,\epsilon)}(\theta_{i})$ is monotonically increasing in terms of  $\log R^{(l)}_{i}$ for our developed FBC-enhanced NOMA schemes; 
	
	\underline{Claim 2:} The uplink $\epsilon$-effective capacity is monotonically increasing in terms of the average power allocation $\overline{\bm{\nu}}^{(l)}_{i}$ for our developed FBC-enhanced NOMA schemes.
\end{theorem}
\begin{IEEEproof}
		The proof of Theorem~\ref{theorem08} is in Appendix D.
\end{IEEEproof}

\indent\textit{Remarks on Theorem~\ref{theorem08}:} According to Theorem~\ref{theorem08}, we can then prove the maximum $\epsilon$-effective energy efficiency that is always achieved at a finite average transmit power, leading to the following theorem.

\begin{theorem}  \label{theorem09}
	Under the constraints $n> \max \left\{\left(n^{\text{rt}}\right)^{2}, n^{\text{th}}\right\}$, an optimal average power allocation policy, denoted by $\overline{\bm{\nu}}_{i}^{(l,\text{opt})}$, for maximizing the uplink $\epsilon$-effective energy efficiency concerning the demands of statistical delay and error-rate bounded QoS is determined as follows:
	\begin{align}\label{equation98}
	\overline{\bm{\nu}}^{(l,\text{opt})}_{i}=&\,\frac{EC^{(l,\epsilon)}(\theta_{i})}{\frac{\partial EC^{(l,\epsilon)}(\theta_{i})}{\partial \overline{\bm{\nu}}^{(l)}_{i}}\Big|_{\overline{\bm{\nu}}^{(l)}_{i}=\overline{\bm{\nu}}^{(l,\text{opt})}_{i}}}-\frac{{\cal P}_{c}}{\eta_{i}},
	\end{align}
	which represents a recursive equation that can be solved through numerical methods.	
\end{theorem}
\begin{IEEEproof}
			The proof of Theorem~\ref{theorem09} is in Appendix E.
\end{IEEEproof}

\indent\textit{Remarks on Theorem~\ref{theorem09}:} Theorem~\ref{theorem09} establishes the optimal average transmit power allocation policy for maximizing the uplink $\epsilon$-effective energy efficiency under the constraints in terms of statistical delay and error-rate bounded QoS for our developed FBC-enhanced NOMA system. The derived policy not only conceptually identifies and quantitatively analyzes the $\epsilon$-effective energy efficiency metric but also provides a set of practically useful techniques for controlling and analyzing energy efficiency performance while stringently upper-bounding both delay and error-rate. This enables a fundamental understanding and insightful analyses of our proposed schemes.
		
\section{Performance Evaluations}\label{sec:results}

\begin{figure*}[!tp]
	\begin{minipage}[t]{0.46\linewidth}
		\centering
		\includegraphics[width=2.8in, height=2.0in]{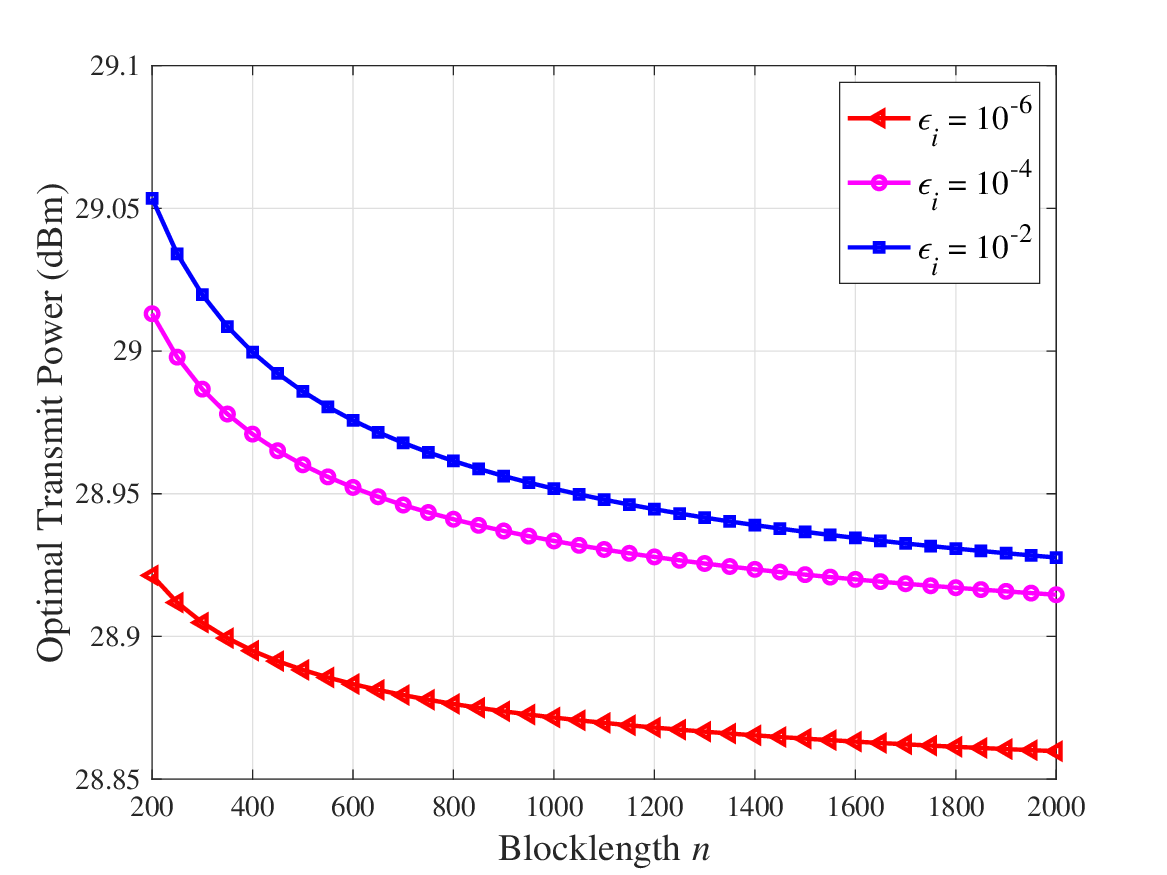}
		\caption{The optimal transmit power vs. blocklength $n$.}
		\label{fig:5}
	\end{minipage}%
	%\hfill
	\hspace{5ex}
	\begin{minipage}[t]{0.46\linewidth}
		\centering
		\includegraphics[width=2.8in, height=2.0in]{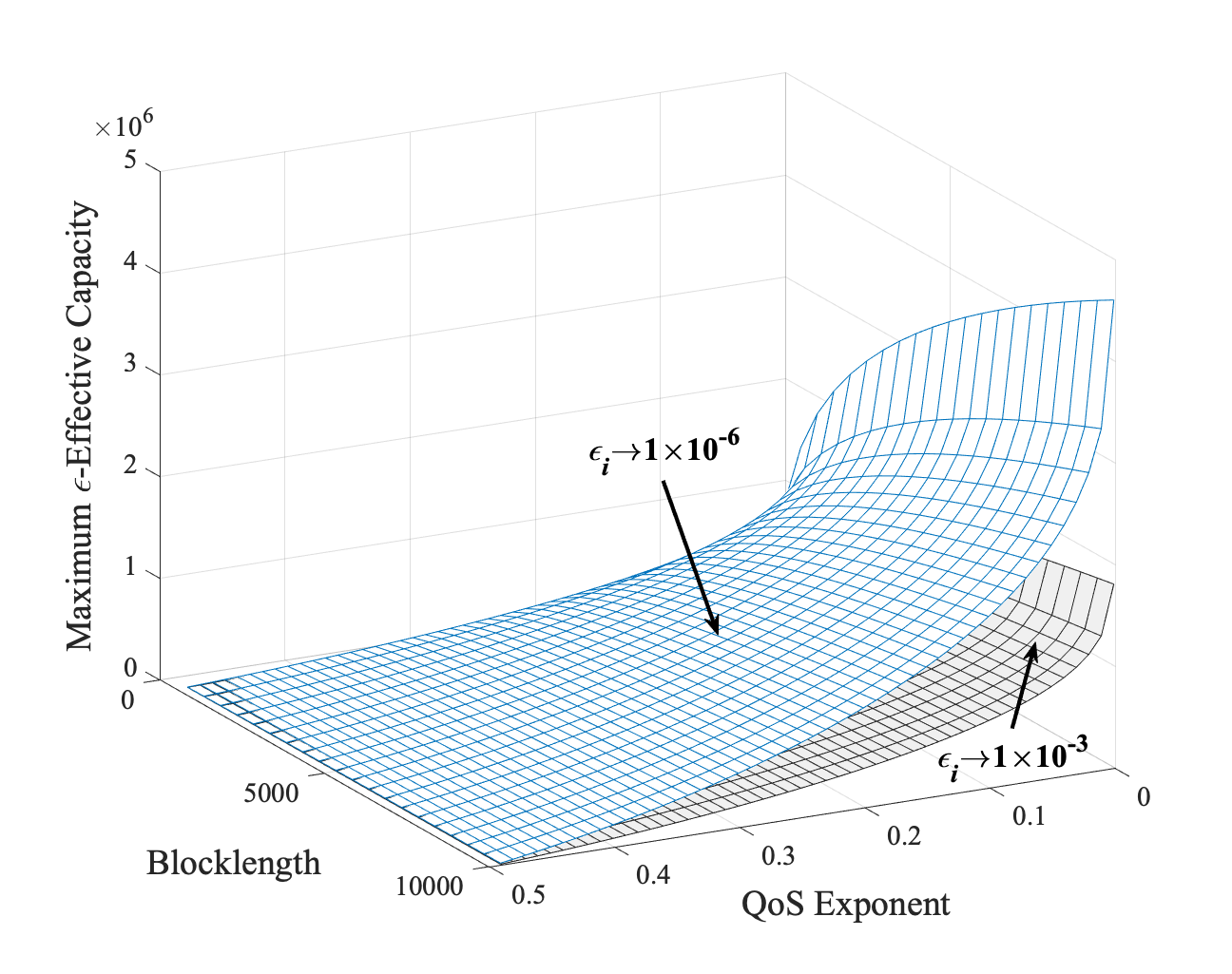}
		\caption{The maximum $\epsilon$-effective capacity vs. blocklength $n$ and QoS exponent $\theta_{i}$ in the finite blocklength regime.}
		\label{fig:3}
	\end{minipage}%
	%\hfill
\end{figure*}

We employ MATLAB to conduct simulations and present a comprehensive set of numerical results, verifying and evaluating the effectiveness of our proposed statistical QoS provisioning schemes.
Throughout our simulations, the average transmit power %vector $\bm{\overline{{\cal P}}}\triangleq\left[\overline{{\cal P}}_{1},\overline{{\cal P}}_{2}\right]$ 
is chosen from $(0, 60]$ dBm for both mobile users and the circuit power ${\cal P}_{c}$ can be chosen from $(0, 200]$ dBm for the BS. In addition, we set the average transmit power efficiency $\eta{_{i}}=\{1,1.2,1.3,1.4\}$.
%The mobile users' delay requirement $T_{\max}$ is set between 50 ms and 100 ms. 

We set the SNR $\gamma^{(l)}_{i}=20$ dB. 
By utilizing the optimal power allocation policy obtained from Theorem~\ref{theorem03}, Fig.~\ref{fig:5} illustrates the optimal transmit power with varying blocklength $n$ in our proposed two-user FBC-enhanced NOMA schemes. It is evident from Fig.~\ref{fig:5} that, for a target decoding error probability, the optimal transmit power exhibits a decreasing trend with respect to $n$.
Furthermore, as depicted in Fig.~\ref{fig:5}, the mobile users are required to assign more power to compensate for the rising decoding error probability.

%Define the normalized effective capacity as the effective capacity divided by $B_{k}$ and $T_{f}$, which then has the unit of bits/sec/Hz. 
%For example, if the required packet loss probability is set to be $10^{-3}$ (i.e., reliability requirement of traditional services), and the  delay is around %1 ms,
Figure~\ref{fig:3} illustrates the maximum $\epsilon$-effective capacity against different blocklengths $n$ and QoS exponents $\theta_{i}$ in our proposed two-user FBC-enhanced NOMA schemes. It is evident from Fig.~\ref{fig:3} that, given a target decoding error probability $\epsilon$, the maximum $\epsilon$-effective capacity decreases as  $\theta_{i}$ increases. 
Additionally, the figure demonstrates that a smaller decoding error probability leads to a higher value of maximum $\epsilon$-effective capacity.

\begin{figure*}[!tp]
	\begin{minipage}[t]{0.46\linewidth}
		\centering
		\includegraphics[width=2.8in, height=2.0in]{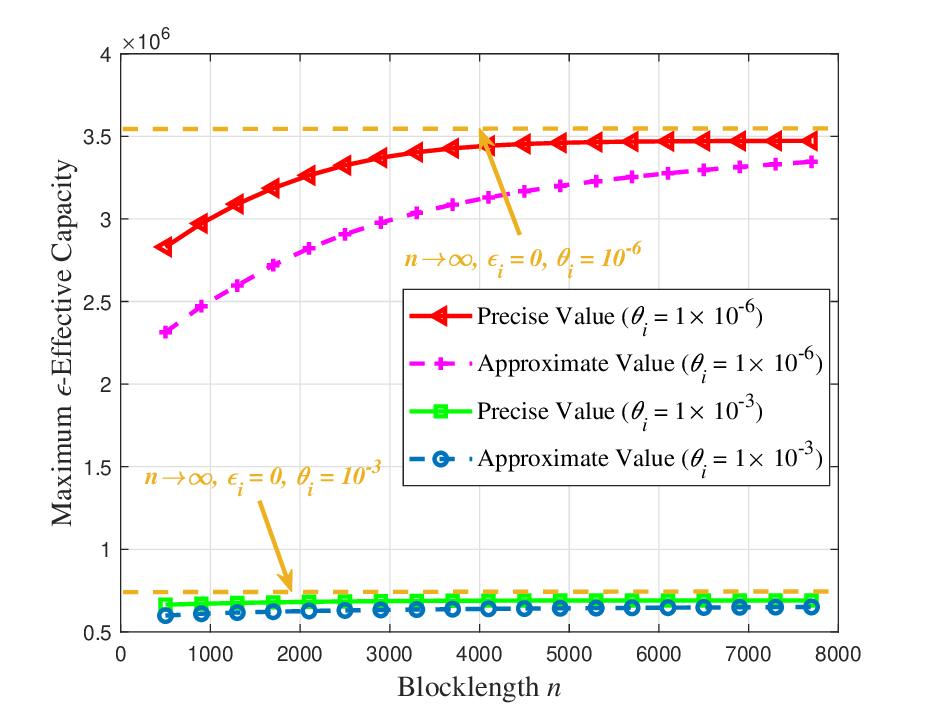}
		\caption{The precise value and approximated value of the maximum $\epsilon$-effective capacity vs. blocklength $n$ when $\theta_{i}=\{1\times10^{-6}, 1\times10^{-3}\}$.}
		\label{fig:2}
	\end{minipage}
	%\hfill
	\hspace{5ex}
	\begin{minipage}[t]{0.46\linewidth}
		\centering
		\includegraphics[width=2.8in, height=2.0in]{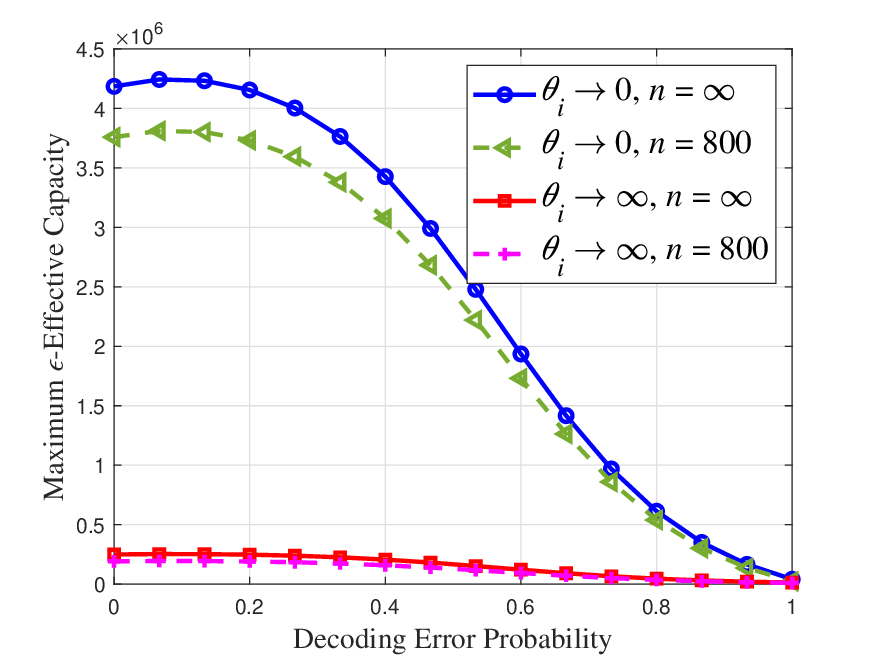}
		\caption{The maximum $\epsilon$-effective capacity vs. decoding error probability $\epsilon$ when $\theta_{i}\rightarrow 0$ and $\theta_{i}\rightarrow \infty$, respectively.}
		\label{fig:4}
	\end{minipage}
\end{figure*}

Now we set the decoding error probability $\epsilon_{i}=10^{-3}$, SNR $\gamma^{(l)}_{i}=20$ dB, transmit power ${\cal P}_{i}=20$ dBm, and $\theta_{i}={1\times10^{-6}, 1\times10^{-3}}$.
We utilize Theorem~\ref{theorem04} to generate Fig.~\ref{fig:2}, which depicts both the precise and approximated values with respect to the maximum $\epsilon$-effective capacity across varied blocklengths $n$ for the developed schemes. 
As illustrated in Fig.~\ref{fig:2}, both the precise and approximated values with respect the maximum $\epsilon$-effective capacity initially rises with the blocklength $n$ before stabilizing at a specific value, aligning with the results from Lemma~\ref{lemma02}. The dashed lines in Fig.~\ref{fig:2} represent the effective capacity for the infinite-blocklength model, where we have $n\rightarrow\infty$ and $\epsilon_{i}=0$, indicating that the transmission rate equals the Shannon capacity. 
Notably, the approximation of the maximum $\epsilon$-effective capacity consistently provides a lower bound on the maximum $\epsilon$-effective capacity.

In Fig.~\ref{fig:4}, the maximum $\epsilon$-effective capacity is plotted against different decoding error probabilities $\epsilon$ for the cases when $\theta_{i}\rightarrow 0$ and $\theta_{i}\rightarrow \infty$. 
As depicted in Fig.~\ref{fig:4}, the maximum $\epsilon$-effective capacity approaches 0 as the decoding error probability $\epsilon_{i}\rightarrow 1$. Additionally, $\theta_{i}\rightarrow 0$ and $\theta_{i}\rightarrow \infty$ serve as the upper and lower bounds on the $\epsilon$-effective capacity, respectively. Notably, Fig.~\ref{fig:4} reveals that as $n\rightarrow\infty$, the conventional effective capacity function in the asymptotic regime is achieved.

\begin{figure*}[!tp]
	\begin{minipage}[t]{0.46\linewidth}
		\centering
		\includegraphics[width=2.8in, height=2.0in]{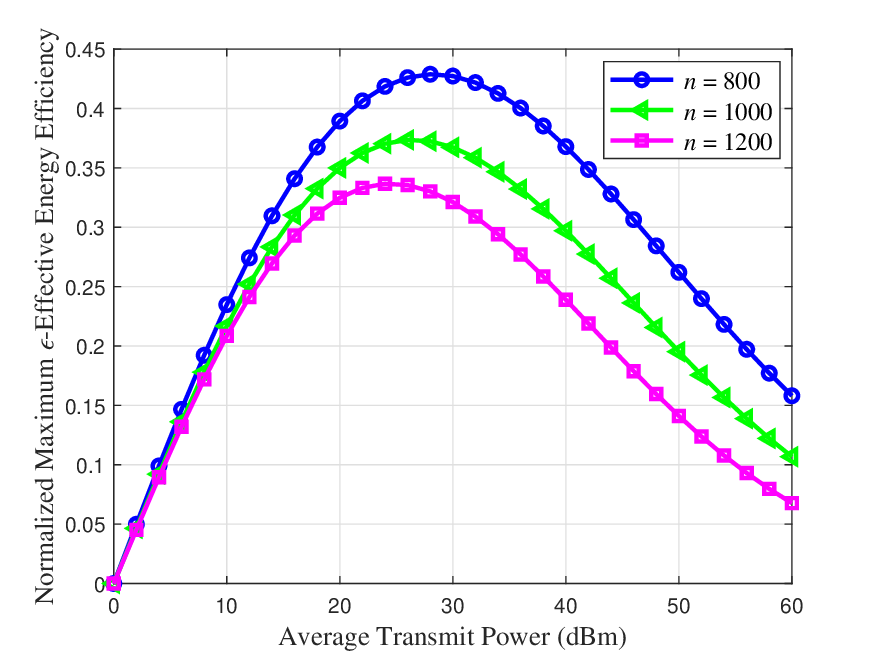}
		\caption{The maximum $\epsilon$-effective energy efficiency vs. average transmit power within the finite blocklength regime.}
		\label{fig:12}
	\end{minipage}%
	%\hfill
	\hspace{5ex}
	\begin{minipage}[t]{0.46\linewidth}
		\centering
		\includegraphics[width=2.8in, height=2.0in]{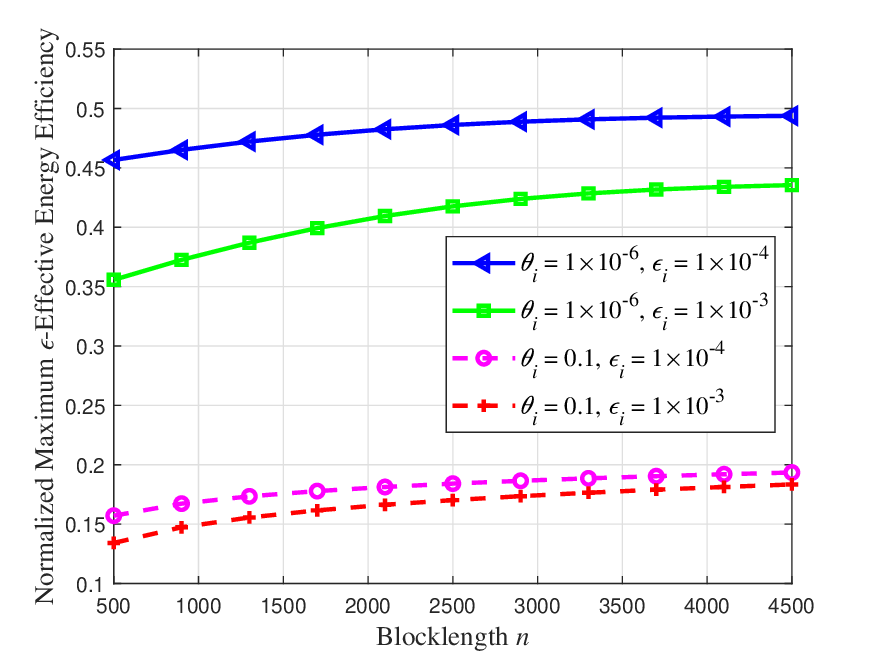}
		\caption{The maximum $\epsilon$-effective energy efficiency vs. blocklength $n$ considering $\theta_{i}=\{1\times10^{-6}, 0.1\}$.}
		\label{fig:14}
	\end{minipage}
\end{figure*}

With the SNR set to $\gamma^{(l)}_{i}=20$ dB, average transmit power efficiency $\eta_{{i}}=1.4$, circuit power ${\cal P}_{c}=40$ dBm, and $\theta_{i}=1\times10^{-6}$, Fig.~\ref{fig:12} illustrates the $\epsilon$-effective energy efficiency plotted against $\overline{{\cal P}}_{i}$ in FBC-enhanced NOMA system. The graph in Fig.~\ref{fig:12} reveals the presence of an optimal average transmit power that maximizes the $\epsilon$-effective energy efficiency, given different blocklengths $n=\{800, 1000, 1200\}$. This observation aligns with the findings presented in Lemma~\ref{lemma05}.

We set the SNR $\gamma^{(l)}_{i}=15$~dB, average transmit power efficiency $\eta{_{i}}=1.3$, circuit power ${\cal P}_{c}=40$ dBm, and $\theta_{i}=\{1\times10^{-6}, 0.1\}$.  
Fig.~\ref{fig:14} depicts the maximum $\epsilon$-effective energy efficiency as a function of $n$. The graph illustrates an initial increase in the maximum $\epsilon$-effective energy efficiency in terms of $n$, followed by a stabilization at a specific value. Notably, Fig.~\ref{fig:14} also indicates that, for a given QoS exponent $\theta_{i}$, the maximum $\epsilon$-effective energy efficiency decreases with higher decoding error probabilities.

\section{Conclusions}\label{sec:conclusion}
We have proposed to formulate and solve the FBC-based $\epsilon$-effective capacity maximization problem for FBC-enhanced NOMA systems, while adhering to statistical constraints on delay and error rates within the framework of QoS provisioning.
Particularly, we have defined system architecture models for uplink two-user FBC-enhanced NOMA while focusing on FBC-based channel coding rates.
In the context of statistical QoS demands, characterized by constraints on both delay and reliability, we have addressed the $\epsilon$-effective capacity maximization problems. Additionally, we have deduced the optimal average power allocation policy aiming to maximize $\epsilon$-effective energy efficiency assuming unknown CSI.
Finally, we have conducted a series of simulations to verify and examine the efficacy of the developed optimal power allocation policies for our proposed two-user FBC-enhanced NOMA schemes.

\begin{appendices}
	\section{Proof for Theorem 1}
To demonstrate the convexity of $\mathbf{P_{2}}$ given by Eq.~(\ref{equation021}), we examine the convexity of $F(n,\epsilon_{2},\gamma^{(1)}_{2})$ within the space spanned by $(n, \bm{\nu}^{(1)}_{2})$.
Subsequently, the Hessian matrix of $F(n,\epsilon_{2},\gamma^{(1)}_{2})$, denoted as $M_{F}$, can be derived as follows:
\begin{equation}\label{equation024}
	M_{F}=
	\left[
	\begin{array}{cccc}
		\frac{\partial^{2} F(n,\epsilon_{2},\gamma^{(1)}_{2})}{\partial n^{2}}&\frac{\partial^{2} F(n,\epsilon_{2},\gamma^{(1)}_{2})}{\partial n\partial \bm{\nu}^{(1)}_{2}}\\
		\frac{\partial^{2} F(n,\epsilon_{2},\gamma^{(1)}_{2})}{\partial \bm{\nu}^{(1)}_{2} \partial n}&\frac{\partial^{2} F(n,\epsilon_{2},\gamma^{(1)}_{2})}{\partial (\bm{\nu}^{(1)}_{2})^{2}}
	\end{array}
	\right].
\end{equation}
Then, the second-order and partial derivative of $F(n,\epsilon_{2},\gamma^{(1)}_{2})$ can be expressed as follows:
\begin{align}\label{equation025}
	\begin{cases}
		\frac{\partial^{2} F(n,\epsilon_{2},\gamma^{(1)}_{2})}{\partial n^{2}}=&\!\!\!\!\sqrt{\left(1-\frac{1}{(1+\bm{\nu}^{(1)}_{2}\gamma^{(1)}_{2})^{2}}\right)} \left[\frac{Q^{-1}\left(\epsilon_{2}\right)\left(n\right)^{-\frac{3}{2}}}{4\log 2}\right]; \\
		\frac{\partial^{2} F(n,\epsilon_{2},\gamma^{(1)}_{2})}{\partial n\partial \bm{\nu}^{(1)}_{2}}=&\!\!\!\!
		\frac{\gamma^{(1)}_{2}}{\left(1+\bm{\nu}^{(1)}_{2}\gamma^{(1)}_{2}\right)\log 2}-\left[\frac{Q^{-1}\left(\epsilon_{2}\right)}{2 \sqrt{n}\log 2}\right] \\
		&\times\frac{\gamma^{(1)}_{2}}{\left(1+\bm{\nu}^{(1)}_{2}\gamma^{(1)}_{2}\right)^{2}\sqrt{\bm{\nu}^{(1)}_{2}\gamma^{(1)}_{2}(\bm{\nu}^{(1)}_{2}\gamma^{(1)}_{2}+2)}};\\
		\frac{\partial^{2} F(n,\epsilon_{2},\gamma^{(1)}_{2})}{\partial \bm{\nu}^{(1)}_{2}\partial n}=&\!\!\!\!
		\frac{\gamma^{(1)}_{2}}{\left(1+\bm{\nu}^{(1)}_{2}\gamma^{(1)}_{2}\right)\log 2}-\left[\frac{Q^{-1}\left(\epsilon_{2}\right)}{2\sqrt{n}\log 2}\right] \\
		&\times\frac{\gamma^{(1)}_{2}}{\left(1+\bm{\nu}^{(1)}_{2}\gamma^{(1)}_{2}\right)^{2}\sqrt{\bm{\nu}^{(1)}_{2}\gamma^{(1)}_{2}(\bm{\nu}^{(1)}_{2}\gamma^{(1)}_{2}+2)}}; \\
		\frac{\partial^{2} F(n,\epsilon_{2},\gamma^{(1)}_{2})}{\partial (\bm{\nu}^{(1)}_{2})^{2}}=&\!\!\!\! -\frac{n\left(\gamma^{(1)}_{2}\right)^{2}}{\left(1+\bm{\nu}^{(1)}_{2}\gamma^{(1)}_{2}\right)^{2}\log 2}+\left[\frac{Q^{-1}\left(\epsilon_{2}\right)\sqrt{n}}{\log 2 }\right]  \\
		&\times \frac{\left(\gamma^{(1)}_{2}\right)^{2}\left(3\left(\bm{\nu}^{(1)}_{2}\gamma^{(1)}_{2}\right)^{2}+6\bm{\nu}^{(1)}_{2}\gamma^{(1)}_{2}+1\right)}{\left(1+\bm{\nu}^{(1)}_{2}\gamma^{(1)}_{2}\right)^{3}\left(\bm{\nu}^{(1)}_{2}\gamma^{(1)}_{2}(\bm{\nu}^{(1)}_{2}\gamma^{(1)}_{2}+2)\right)^{\frac{3}{2}}}.
	\end{cases}
\end{align}
Therefore, we have
\begin{align}\label{equation026}
	&\frac{\partial^{2} F(n,\epsilon_{2},\gamma^{(1)}_{2})}{\partial n^{2}}\frac{\partial^{2} F(n,\epsilon_{2},\gamma^{(1)}_{2})}{\partial (\bm{\nu}^{(1)}_{2})^{2}} \nonumber \\
	&\quad\!\!=\! \frac{\left[Q^{-1}\!\left(\epsilon_{2}\right)\right]^{2}\!\left(\!\gamma^{(1)}_{2}\!\right)^{2}\!\left(3\!\left(\bm{\nu}^{(1)}_{2}\gamma^{(1)}_{2}\right)^{2}\!+\!6\bm{\nu}^{(1)}_{2}\gamma^{(1)}_{2}\!+\!1\right)}{4(\log 2)^{2}n\left(1+\bm{\nu}^{(1)}_{2}\gamma^{(1)}_{2}\right)^{4}\bm{\nu}^{(1)}_{2}\gamma^{(1)}_{2}(\bm{\nu}^{(1)}_{2}\gamma^{(1)}_{2}+2)}\nonumber\\
	&\qquad-\frac{Q^{-1}\left(\epsilon_{2}\right)\left(\gamma^{(1)}_{2}\right)^{2}\sqrt{\bm{\nu}^{(1)}_{2}\gamma^{(1)}_{2}(\bm{\nu}^{(1)}_{2}\gamma^{(1)}_{2}+2)}}{4(\log 2)^{2}\sqrt{n}\left(1+\bm{\nu}^{(1)}_{2}\gamma^{(1)}_{2}\right)^{3}},
\end{align}
and
\begin{align}\label{equation027}
	&\frac{\partial^{2} F(n,\epsilon_{2},\gamma^{(1)}_{2})}{\partial n\partial \bm{\nu}^{(1)}_{2}}\frac{\partial^{2} F(v)}{\partial \bm{\nu}^{(1)}_{2} \partial n} 
	=\left[\frac{\gamma^{(1)}_{2}}{\left(1+\bm{\nu}^{(1)}_{2}\gamma^{(1)}_{2}\right)\log 2}\right]^{2}
	\nonumber\\
	&\qquad+\frac{\left[Q^{-1}\left(\epsilon_{2}\right)\right]^{2}\left(\gamma^{(1)}_{2}\right)^{2}}{4n(\log 2)^{2}\left(1+\bm{\nu}^{(1)}_{2}\gamma^{(1)}_{2}\right)^{4}{\bm{\nu}^{(1)}_{2}\gamma^{(1)}_{2}(\bm{\nu}^{(1)}_{2}\gamma^{(1)}_{2}+2)}} \nonumber \\
	&\qquad-\frac{Q^{-1}\left(\epsilon_{2}\right)\left(\gamma^{(1)}_{2}\right)^{2}}{\sqrt{n}(\log 2)^{2}\left(1+\bm{\nu}^{(1)}_{2}\gamma^{(1)}_{2}\right)^{3}\sqrt{\bm{\nu}^{(1)}_{2}\gamma^{(1)}_{2}(\bm{\nu}^{(1)}_{2}\gamma^{(1)}_{2}+2)}}.
\end{align}
Correspondingly, using Eq.~(\ref{equation026}) and~(\ref{equation027}), we get
\begin{align}\label{equation028}
	&\!\frac{\partial^{2} F(n,\epsilon_{2},\gamma^{(1)}_{2})}{\partial n^{2}}\frac{\partial^{2} F(n,\epsilon_{2},\gamma^{(1)}_{2})}{\partial (\bm{\nu}^{(1)}_{2})^{2}}\nonumber\\
	&\,\,\,-\!\frac{\partial^{2} F(n,\epsilon_{2},\gamma^{(1)}_{2})}{\partial n\partial \bm{\nu}^{(1)}_{2}}\frac{\partial^{2} F(n,\epsilon_{2},\gamma^{(1)}_{2})}{\partial \bm{\nu}^{(1)}_{2} \partial n}\nonumber\\
	&\quad=\!\frac{\!\left(\gamma^{(1)}_{2}\right)^{2}}{4n(\log 2)^{2}}\vast\{\!\frac{\left[Q^{-1}\!\left(\!\epsilon_{2}\!\right)\right]^{\!2}\!\left(\!3\!\left(\!\bm{\nu}^{(1)}_{2}\gamma^{(1)}_{2}\!\right)^{2}\!\!+\!6\bm{\nu}^{(1)}_{2}\gamma^{(1)}_{2}\!+\!1\right)}{\left(1+\bm{\nu}^{(1)}_{2}\gamma^{(1)}_{2}\right)^{4}\bm{\nu}^{(1)}_{2}\gamma^{(1)}_{2}\left(\bm{\nu}^{(1)}_{2}\gamma^{(1)}_{2}+2\right)} \nonumber\\
	&\qquad\!+\!\frac{4\sqrt{n}Q^{-1}\left(\epsilon_{2}\right)}{\left(1+\bm{\nu}^{(1)}_{2}\gamma^{(1)}_{2}\right)^{3}\sqrt{\bm{\nu}^{(1)}_{2}\gamma^{(1)}_{2}\left(\bm{\nu}^{(1)}_{2}\gamma^{(1)}_{2}+2\right)}}\nonumber\\
	&\qquad\!-\!\frac{4n}{\!\left(\!1\!+\!\bm{\nu}^{(1)}_{2}\gamma^{(1)}_{2}\!\right)^{\!2}}
	\!-\!\frac{\left[Q^{-1}\left(\epsilon_{2}\right)\right]^{2}}{\left(\!1\!+\!\bm{\nu}^{(1)}_{2}\gamma^{(1)}_{2}\!\right)^{\!4}\bm{\nu}^{(1)}_{2}\gamma^{(1)}_{2}\!\left(\bm{\nu}^{(1)}_{2}\gamma^{(1)}_{2}\!+\!2\right)} \nonumber\\
	&\qquad\!-\!\frac{\sqrt{n}Q^{-1}\left(\epsilon_{2}\right)\sqrt{\bm{\nu}^{(1)}_{2}\gamma^{(1)}_{2}\left(\bm{\nu}^{(1)}_{2}\gamma^{(1)}_{2}+2\right)}}{\left(1+\bm{\nu}^{(1)}_{2}\gamma^{(1)}_{2}\right)^{3}}\vast\}.
\end{align}
By employing Eq.~(\ref{equation028}), it is sufficient to have
\begin{align}\label{equation029}
	&\frac{3\left[Q^{-1}\left(\epsilon_{2}\right)\right]^{2}}{\left(1\!+\!\bm{\nu}^{(1)}_{2}\gamma^{(1)}_{2}\right)^{4}}
	\!+\!\frac{4\sqrt{n}Q^{-1}\left(\epsilon_{2}\right)}{\left(1\!+\!\bm{\nu}^{(1)}_{2}\gamma^{(1)}_{2}\right)^{3}\sqrt{\bm{\nu}^{(1)}_{2}\gamma^{(1)}_{2}\left(\bm{\nu}^{(1)}_{2}\gamma^{(1)}_{2}\!+\!2\right)}}\nonumber\\
	&\quad\!-\!\frac{4n}{\left(\!1\!+\!\bm{\nu}^{(1)}_{2}\gamma^{(1)}_{2}\!\right)^{2}} 
	\!-\!\frac{\sqrt{n}Q^{-1}\!\!\left(\epsilon_{2}\right)\!\!\sqrt{\bm{\nu}^{(1)}_{2}\gamma^{(1)}_{2}\!\left(\!\bm{\nu}^{(1)}_{2}\gamma^{(1)}_{2}\!+\!2\right)}}{\left(1+\bm{\nu}^{(1)}_{2}\gamma^{(1)}_{2}\right)^{3}} 
	\nonumber\\
	&\qquad>0,
\end{align}
for guaranteeing the following inequality:
\begin{align}\label{equation018}
	&\frac{\partial^{2} \!F(n,\epsilon_{2},\gamma^{(1)}_{2})}{\partial n^{2}}\frac{\partial^{2}\! F(n,\epsilon_{2},\gamma^{(1)}_{2})}{\partial (\bm{\nu}^{(1)}_{2})^{2}}
	\nonumber \\
	&\qquad >\!\frac{\partial^{2}\! F(n,\epsilon_{2},\gamma^{(1)}_{2})}{\partial n\partial \bm{\nu}^{(1)}_{2}}\frac{\partial^{2}\! F(n,\epsilon_{2},\gamma^{(1)}_{2})}{\partial \bm{\nu}^{(1)}_{2} \partial n}.
\end{align}
Thus, we have
\begin{align}\label{equation290}
	&-4n
	+\frac{4\sqrt{n}Q^{-1}\left(\epsilon_{2}\right)}{\left(1+\bm{\nu}^{(1)}_{2}\gamma^{(1)}_{2}\right)\sqrt{\bm{\nu}^{(1)}_{2}\gamma^{(1)}_{2}\left(\bm{\nu}^{(1)}_{2}\gamma^{(1)}_{2}+2\right)}}  \nonumber\\
	&\qquad-\frac{\sqrt{n}Q^{-1}\left(\epsilon_{2}\right)\sqrt{\bm{\nu}^{(1)}_{2}\gamma^{(1)}_{2}\left(\bm{\nu}^{(1)}_{2}\gamma^{(1)}_{2}\!+\!2\right)}}{\left(1+\bm{\nu}^{(1)}_{2}\gamma^{(1)}_{2}\right)}\nonumber\\
	&\qquad+\frac{3\left[Q^{-1}\left(\epsilon_{2}\right)\right]^{2}}{\left(1+\bm{\nu}^{(1)}_{2}\gamma^{(1)}_{2}\right)^{2}}\!>\!0.
\end{align}
The left-hand side of Eq.~(\ref{equation290}) is a quadratic equation of $\sqrt{n}$. Setting $\overline{a}= 1+\bm{\nu}^{(1)}_{2}\gamma^{(1)}_{2}$ and $\overline{b}=\sqrt{\bm{\nu}^{(1)}_{2}\gamma^{(1)}_{2}\left(\bm{\nu}^{(1)}_{2}\gamma^{(1)}_{2}+2\right)}$, 
we can rewrite Eq.~(\ref{equation290}) as in the following equation:
\begin{align}\label{equation030}
	&-4n\!+\!4\frac{\sqrt{n}Q^{-1}\!\left(\epsilon_{2}\right)}{\overline{a}\overline{b}}-\frac{\sqrt{n}Q^{-1}\!\left(\epsilon_{2}\right)\overline{b}}{\overline{a}} \!+\!\frac{3\!\left[Q^{-1}\!\left(\epsilon_{2}\right)\right]^{2}}{(\overline{a})^{2}}\nonumber\\
	&\qquad>0.
\end{align}
Accordingly, when $\epsilon_{2}\in(0,1/2)$, the positive root, denoted by $n^{\text{rt}}$, of Eq.~(\ref{equation030}) is expressed as follows:
\begin{align}\label{equation031}
	n^{\text{rt}}=&\left[\frac{Q^{-1}\left(\epsilon_{2}\right)}{8}\right]\left(\frac{4}{\overline{a}\overline{b}}-\frac{\overline{b}}{\overline{a}}\right)+\left[{\frac{Q^{-1}\left(\epsilon_{2}\right)}{2}}\right] 
	\nonumber\\
	&\times \sqrt{\frac{1}{16}\left(\frac{4}{\overline{a}\overline{b}}-\frac{\overline{b}}{\overline{a}}
		\right)^{2}+3\left[Q^{-1}\left(\epsilon_{2}\right)\right]^{2}}.
\end{align}
Using Eqs.~(\ref{equation018}) and~(\ref{equation031}), we can prove that 
\begin{equation}
	\begin{cases}
		\frac{\partial^{2} F(n,\epsilon_{2},\gamma^{(1)}_{2})}{\partial n^{2}}>0;  \\
		\frac{\partial^{2} F(n,\epsilon_{2},\gamma^{(1)}_{2})}{\partial n^{2}}\frac{\partial^{2} F(n,\epsilon_{2},\gamma^{(1)}_{2})}{\partial (\bm{\nu}^{(1)}_{2})^{2}}\\
		\qquad\qquad-\frac{\partial^{2} F(n,\epsilon_{2},\gamma^{(1)}_{2})}{\partial n\partial \bm{\nu}^{(1)}_{2}}\frac{\partial^{2} F(n,\epsilon_{2},\gamma^{(1)}_{2})}{\partial \bm{\nu}^{(1)}_{2} \partial n}>0,
	\end{cases}
\end{equation}
given the following constraint:
\begin{equation}\label{equation052}
	n> \max \left\{\left(n^{\text{rt}}\right)^{2}, n^{\text{th}}\right\}.
\end{equation}
Therefore, through applying Sylvester's criterion, by demonstrating that $F(n,\epsilon_{2},\gamma^{(1)}_{2})$ is a convex function within the space defined by $(n, \bm{\nu}^{(1)}_{2})$ under the conditions $\epsilon_{2}\in(0,1/2)$ and $n> \max \left\{\left(n^{\text{rt}}\right)^{2}, n^{\text{th}}\right\}$, we establish the convexity of $\mathbf{P_{2}}$. 
By employing Eq.~(\ref{equation021}), the auxiliary function $G(F(n,\epsilon_{2},\gamma_{2}))$ is defined as the objective function of the optimization problem $\mathbf{P_{2}}$ presented in Eq.~(\ref{equation021}), which is expressed as follows:
\begin{equation}
	G(F(n,\epsilon_{2},\gamma^{(1)}_{2}))\triangleq \frac{-\theta_{2}F(n,\epsilon_{2},\gamma^{(1)}_{2})}{n}.
\end{equation}
For characterizing the convexity of $G(F(n,\epsilon_{2},\gamma^{(1)}_{2}))$, the composition rules of convex functions is applied to prove the following equation:
\begin{align}
	\frac{\partial^{2} G(F(n,\epsilon_{2},\gamma^{(1)}_{2}))}{\partial n^{2}}>0.
\end{align}
Then, we establish that the objective function of $\mathbf{P_{2}}$  defined in Eq.~(\ref{equation021}) is strictly convex within the space defined by $(n, \bm{\nu}^{(1)}_{2})$. This holds true under the conditions $0<\epsilon_{2}<1/2$ and $n> \max \left\{\left(n^{\text{rt}}\right)^{2}, n^{\text{th}}\right\}$. This completes the proof of Theorem~\ref{theorem02}.

\section{Proof for Theorem 2}
		
Based on the results derived by Theorem~\ref{theorem02}, we substitute constraints (b) and (c) specified in Eq.~(\ref{equation019}) into Eq.~(\ref{equation020}).
Subsequently, it is sufficient to focus on the average power constraint (a) outlined by Eq.~(\ref{equation019}) for obtaining a closed-form solution to $\mathbf{P_{2}}$ as defined by Eq.~(\ref{equation021}). 
Initially, the Lagrange function, denoted as $J$, can be formulated as follows:
\begin{align}\label{equation032}
	J\!=&\,\mathbb{E}_{\gamma^{(1)}_{2}}\!\Bigg[\epsilon_{2}\!+\!(1\!-\!\epsilon_{2})\exp\!\Bigg(\!\!-\theta_{2}\!\Bigg(\log_{2}\!\left(1\!+\!\bm{\nu}^{(1)}_{2}\gamma^{(1)}_{2}\right)\nonumber\\
	&\quad-\!  \left[\!\frac{Q^{-1}\left(\epsilon_{2}\right)}{\log 2}\!\right]\sqrt{\frac{1}{n}\left(1\!-\!\frac{1}{(1+\bm{\nu}^{(1)}_{2}\gamma^{(1)}_{2})^{2}}\!\right)\!}\Bigg)\Bigg)\Bigg]\nonumber\\
	&\quad+\!\widetilde{\lambda}_{2}\left(\mathbb{E}_{\gamma^{(1)}_{2}}\left[{\cal P}_{2}(\bm{\nu}^{(1)})\right]\!-\!\overline{{\cal P}}_{2}\right) \nonumber\\
	=&\,\mathbb{E}_{\gamma^{(1)}_{2}}\!\vvast[\epsilon_{2}\!+\!(1\!-\!\epsilon_{2})\exp\!\vvast(\!\!\sqrt{\!\frac{1}{n}\!\left(\!1\!-\!\frac{1}{(1\!+\!\bm{\nu}^{(1)}_{2}\gamma^{(1)}_{2})^{2}}\!\right)}\nonumber\\
	&\quad\times\left[\frac{\theta_{2}Q^{-1}\!\left(\epsilon_{2}\right)}{\log 2}\!\right]\!\vvast)\left(1+\bm{\nu}^{(1)}_{2}\gamma^{(1)}_{2}\!\right)^{-\beta_{2}}\!\vvast] \nonumber\\
	&\quad +\widetilde{\lambda}_{2}\left(\mathbb{E}_{\gamma^{(1)}_{2}}\left[{\cal P}_{2}(\bm{\nu}^{(1)})\right]-\overline{{\cal P}}_{2}\right)
\end{align}
where $\widetilde{\lambda}_{2}$ denotes the Lagrange multiplier corresponding to constraint (a) as specified by Eq.~(\ref{equation019}). In high SNR region, i.e., $\bm{\nu}^{(1)}_{2}\gamma^{(1)}_{2}\gg1$, we can rewrite the Lagrange function as follows:
\begin{align}\label{equation033}
	J=&\,\mathbb{E}_{\gamma^{(1)}_{2}}\left[\epsilon_{2}+(1-\epsilon_{2})e^{\frac{\theta_{2}Q^{-1}\left(\epsilon_{2}\right)}{\sqrt{n}\log 2}}\left(\bm{\nu}^{(1)}_{2}\gamma^{(1)}_{2}\right)^{-\beta_{2}}\right]+\widetilde{\lambda}_{2}
	\nonumber\\
	&\times \left(\mathbb{E}_{\gamma^{(1)}_{2}}\left[{\cal P}_{2}(\bm{\nu}^{(1)})\right]-\overline{{\cal P}}_{2}\right).
\end{align}
Subsequently, by implementing the Karush-Kuhn-Tucker (KKT) condition and taking the first derivative of $J$ in terms of $\bm{\nu}^{(1)}_{2}$ and equate the results to zero, it yields:
\begin{align}\label{equation034}
	\frac{\partial J}{\partial\bm{\nu}^{(1)}_{2}}=&\!-(1\!-\!\epsilon_{2})e^{\frac{\theta_{2}Q^{-1}\left(\epsilon_{2}\right)}{\sqrt{n}\log 2}} \beta_{2}\gamma^{(1)}_{2}\!\left(\bm{\nu}^{(1)}_{2}\gamma^{(1)}_{2}\right)^{-\beta_{2}-1} \!\!\!\!\!\!p_{\Gamma}(\gamma^{(1)}_{2})\nonumber\\
	&+\widetilde{\lambda}_{2}p_{\Gamma}(\gamma^{(1)}_{2}) 
	=0.
\end{align}
Consequently, using Eq.~(\ref{equation034}), in the presence of the constraints of statistical delay and error-rate in high SNR region, the \textit{optimal power allocation policy} is developed for maximizing the uplink $\epsilon$-effective capacity, as detailed by Eq.~\eqref{equation035a}, concluding the proof of Theorem~\ref{theorem03}.

\section{Proof for Theorem 3}
First, the following two auxiliary functions are defined:
\begin{equation}
	\begin{cases}
		f(\gamma^{(l)}_{i})\triangleq\exp\left(
		\left[\frac{\theta_{i}Q^{-1}(\epsilon_{i})}{\sqrt{n}\log 2}\right] \sqrt{1-\frac{1}{\left(\widetilde{a}\beta_{i}\gamma^{(l)}_{i}\right)^{2}}}\right); \\
		g(\gamma^{(l)}_{i})\triangleq\left(\widetilde{a}\beta_{i}\gamma^{(l)}_{i}\right)^{-\frac{\beta_{i}}{\beta_{i}+1}}.
	\end{cases}
\end{equation}
Since $ \mathbb{E}_{\gamma^{(l)}_{i}}[f(\gamma^{(l)}_{i})g(\gamma^{(l)}_{i})]>\mathbb{E}_{\gamma^{(l)}_{i}}[f(\gamma^{(l)}_{i})]\mathbb{E}_{\gamma^{(l)}_{i}}[g(\gamma^{(l)}_{i})]$, the approximate maximum uplink $\epsilon$-effective capacity $EC^{(l,\epsilon)}_{\text{max}}(\theta_{i})$ $(i\in\{1,2\})$ is determined as in the following equations:
\begin{align}\label{equation039}
	&EC^{(l,\epsilon)}_{\text{max}}(\theta_{i})\nonumber\\
	&\!\quad\!	\approx\!-\frac{1}{\theta_{i}}\log\left(1-\epsilon_{i}\right)-\frac{1}{\theta_{i}}\log\left\{\mathbb{E}_{\gamma^{(l)}_{i}}\left[
	\left(\widetilde{a}\beta_{i}\gamma^{(l)}_{i}\right)^{-\frac{\beta_{i}}{\beta_{i}+1}}\right]\right\}\nonumber\\
	&\quad\!\quad\!
	-\frac{1}{\theta_{i}}\log\Bigg\{\mathbb{E}_{\gamma^{(l)}_{i}}\!\Bigg[\exp\!\Bigg(
	\left[\frac{\theta_{i}Q^{-1}(\epsilon_{i})}{\sqrt{n}\log 2}\right]
	\nonumber\\
	&\quad\!\quad\!\times  \!\sqrt{\left(1\!-\left(\widetilde{a}\beta_{i}\gamma^{(l)}_{i}\right)^{-\frac{2}{\beta_{i}+1}}\right)}\Bigg)\!\Bigg]\!\Bigg\}
	\nonumber \\
	%	&	=-\frac{1}{\theta_{i}}\log\left(1-\epsilon_{i}\right)-\frac{1}{\theta_{i}}\log\left\{\int_{0}^{\infty}	\left(\widetilde{a}\beta_{i}\gamma^{(l)}_{i}\right)^{-\frac{\beta_{i}}{\beta_{i}+1}}p_{\Gamma}(\gamma^{(l)}_{i})d\gamma^{(l)}_{i}\right\}\nonumber\\
	%	&\quad
	%	-\frac{1}{\theta_{i}}\log\Bigg\{\int_{0}^{\infty}\exp\left(\left[\frac{\theta_{i}Q^{-1}(\epsilon_{i})}{\sqrt{n}\log 2}\right] \sqrt{\left(1-\left(\widetilde{a}\beta_{i}\gamma^{(l)}_{i}\right)^{-\frac{2}{\beta_{i}+1}}\right)}\right) p_{\Gamma}(\gamma^{(l)}_{i})d\gamma^{(l)}_{i}\Bigg\}
	%	\nonumber \\
	&\quad\!
	=\!-\frac{1}{\theta_{i}}\!\Bigg\{\!\!\log\!\left(1\!-\!\epsilon_{i}\right)\!+\!\log\!\left\{\! \left(\widetilde{a}\beta_{i}\overline{\gamma}_{i}\right)^{-\frac{\beta_{i}}{\beta_{i}+1}} \Gamma\!\left(\frac{1}{\beta_{i}\!+\!1},\frac{\widetilde{\lambda}^{\text{opt}}_{i}}{\beta_{i}}\!\right)\!\right\} \nonumber\\
	&\quad\!\quad\!
	+\log\!\Bigg\{\!\!\int_{0}^{\infty}\!\!\exp\!\Bigg(\! \left[\frac{\theta_{i}Q^{-1}(\epsilon_{i})}{\!\!\sqrt{n}\log 2}\right] \sqrt{\!\left(\!1\!-\!\left(\widetilde{a}\beta_{i}\gamma^{(l)}_{i}\right)^{-\frac{2}{\beta_{i}+1}}\!\right)\!}\Bigg) \nonumber\\
	&\quad\!\quad\!\times  p_{\Gamma}(\gamma^{(l)}_{i})d\gamma^{(l)}_{i}\Bigg\}\Bigg\}
	\nonumber \\
	&\quad\!
	\overset{(a)}{\approx}-\!\frac{1}{\theta_{i}}\!\Bigg\{\!\!\log\left(1\!-\!\epsilon_{i}\right)\!+\!\log\!\left\{\!
	\!\left(\widetilde{a}\beta_{i}\overline{\gamma}_{i}\right)^{-\frac{\beta_{i}}{\beta_{i}+1}}\Gamma\!\left(\frac{1}{\beta_{i}\!+\!1},\!\frac{\widetilde{\lambda}^{\text{opt}}_{i}}{\beta_{i}}\!\right)\!\!\right\}
	\nonumber\\
	&\quad\quad\!
	+\!\!\int_{0}^{\infty}\!
	\left[\frac{\theta_{i}Q^{-1}(\epsilon_{i})}{\sqrt{n}\log 2}\right]\!\! \sqrt{\!1\!-\!\left(\widetilde{a}\beta_{i}\gamma^{(l)}_{i}\!\right)^{\!-\frac{2}{\beta_{i}+1}}}	 p_{\Gamma}(\gamma^{(l)}_{i})d\gamma^{(l)}_{i}\!\Bigg\},
\end{align}
where $(a)$ is derived from applying Jensen's inequality.
Furthermore, to derive the term 
$\sqrt{\left(1-\left(\widetilde{a}\beta_{i}\gamma^{(l)}_{i}\right)^{-\frac{2}{\beta_{i}+1}}\right)}$ in the integral at the right-hand side of (a) in Eq.~(\ref{equation039}),  we define
\begin{equation}
	\widetilde{x}\triangleq \left(\widetilde{a}\beta_{i}\gamma^{(l)}_{i}\right)^{-\frac{2}{\beta_{i}+1}}.
\end{equation}  
Subsequently, through applying Taylor expansion with respect to the function $\sqrt{(1-\widetilde{x})}$ considering $\widetilde{x}\cong 0$, %and cutting off at  $\widetilde{x}^{3}$
the term  $\sqrt{\left(1-\left(\widetilde{a}\beta_{i}\gamma^{(l)}_{i}\right)^{-\frac{2}{\beta_{i}+1}}\right)}$ inside the integral at the right-hand side of (a) in Eq.~(\ref{equation039}) can be expressed as follows: 
\begin{align}
	&\sqrt{\!\left(\!1\!-\!\left(\widetilde{a}\beta_{i}\gamma^{(l)}_{i}\right)^{-\frac{2}{\beta_{i}+1}}\!\right)}	\nonumber\\
	&\quad=\,1\!-\!\frac{\left(\widetilde{a}\beta_{i}\gamma^{(l)}_{i}\right)^{-\frac{2}{\beta_{i}+1}}}{2}-\frac{\left(\widetilde{a}\beta_{i}\gamma^{(l)}_{i}\right)^{-\frac{4}{\beta_{i}+1}}}{8}
	\nonumber\\
	&\qquad-\!\sum_{\ell=3}^{\infty}\frac{(2\ell-3)!}{\ell!(\ell-2)!2^{2\ell\!-\!2}}
	\left(\widetilde{a}\beta_{i}\gamma^{(l)}_{i}\right)^{-\frac{2\ell}{\beta_{i}+1}}.
\end{align}
Correspondingly, we can obtain
\begin{align}\label{equation043}
	&	EC^{(l,\epsilon)}_{\text{max}}(\theta_{i}) \nonumber\\
	&\quad\!	\approx-\frac{1}{\theta_{i}}\Bigg\{\log\Bigg\{\left(1-\epsilon_{i}\right)
	\left(\widetilde{a}\beta_{i}\overline{\gamma}_{i}\right)^{-\frac{\beta_{i}}{\beta_{i}+1}}\Gamma\left(\frac{1}{\beta_{i}+1},\frac{\widetilde{\lambda}^{\text{opt}}_{i}}{\beta_{i}}\right)\!\Bigg\}	\nonumber\\
	&\qquad \!\!	+\int_{0}^{\infty} \left[\frac{\theta_{i}Q^{-1}(\epsilon_{i})}{\overline{\gamma}_{i}\sqrt{n}\log 2}\right] \Bigg[1\!-\!\frac{\left(\widetilde{a}\beta_{i}\gamma^{(l)}_{i}\right)^{-\frac{2}{\beta_{i}+1}}}{2}
	\nonumber\\
	&\qquad\!\!-\!\frac{\left(\widetilde{a}\beta_{i}\gamma^{(l)}_{i}\right)^{-\frac{4}{\beta_{i}+1}}}{8}
	\!-\!\sum_{\ell=3}^{\infty}\frac{(2\ell-3)!}{\ell!(\ell-2)!2^{2\ell-2}}\!
	\left(\widetilde{a}\beta_{i}\gamma^{(l)}_{i}\right)^{-\frac{2\ell}{\beta_{i}+1}}\!\!\Bigg] 	\nonumber\\
	&\qquad\!\!\times\! e^{-\frac{\gamma^{(l)}_{i}}{\overline{\gamma}_{i}}}d\gamma^{(l)}_{i}\Bigg\} 
	%	\nonumber \\
	%	=&\!-\!\frac{1}{\theta_{i}}\Bigg\{\!\log\left\{\!\left(1\!-\!\epsilon_{i}\right)	\left(\widetilde{a}\beta_{i}\overline{\gamma}_{i}\right)^{-\frac{\beta_{i}}{\beta_{i}+1}}\Gamma\left(\frac{1}{\beta_{i}+1},\frac{\widetilde{\lambda}^{\text{opt}}_{i}}{\beta_{i}}\right)\!\right\}\!+\!\left[\frac{\theta_{i}Q^{-1}(\epsilon_{i})}{\sqrt{n}\log 2}\right]\!	\Bigg[1\!-\!\frac{\left(\widetilde{a}\beta_{i}\overline{\gamma}_{i}\right)^{-\frac{2}{\beta_{i}+1}}}{2} \nonumber\\
	%	&\times \Gamma\left(\frac{\beta_{i}-1}{\beta_{i}+1},\frac{\widetilde{\lambda}^{\text{opt}}_{i}}{\beta_{i}}\right)- \frac{\left(\widetilde{a}\beta_{i}\overline{\gamma}_{i}\right)^{-\frac{4}{\beta_{i}+1}}}{8} 	\Gamma\left(\frac{\beta_{i}-3}{\beta_{i}+1},\frac{\widetilde{\lambda}^{\text{opt}}_{i}}{\beta_{i}}\right)	-\sum_{\ell=3}^{\infty}\frac{(2\ell-3)!}{\ell!(\ell-2)!2^{2\ell-2}} \nonumber\\
	%	&\times \left(\widetilde{a}\beta_{i}\overline{\gamma}_{i}\right)^{-\frac{2\ell}{\beta_{i}+1}} 	\Gamma\left(\frac{\beta_{i}-2\ell+1}{\beta_{i}+1},\frac{\widetilde{\lambda}^{\text{opt}}_{i}}{\beta_{i}}\right)\Bigg]\Bigg\},
\end{align}
which is Eq.~(\ref{equation036a}). Therefore, the proof of Theorem~\ref{theorem04} is concluded.

\section{Proof for Theorem 4}
		
We proceed with this proof by showing Theorem~\ref{theorem08}'s \underline{Claim 1} and \underline{Claim 2}, respectively, as follows.

\underline{Claim 1.} Based on Eq.~(\ref{equation017}), the first-order derivative of the $\epsilon$-effective capacity in terms of $\log R^{(l)}_{i}$ is derived through the following equation:
\begin{align}
	\frac{\partial EC^{(l,\epsilon)}(\theta_{i})}{\partial \log R^{(l)}_{i}}=\mathbb{E}_{\gamma^{(l)}_{i}}\left\{\frac{\epsilon_{i}+(1-\epsilon_{i})e^{-\theta_{i}\frac{\log R^{(l)}_{i}}{n}}}{n\mathbb{E}_{\gamma^{(l)}_{i}}\left[\epsilon_{i}+(1-\epsilon_{i})e^{-\theta_{i}\frac{\log R^{(l)}_{i}}{n}}\right]}\right\}>0,
\end{align}
which shows that $EC^{(l,\epsilon)}(\theta_{i})$ is a monotonically increasing function of $\log R^{(l)}_{i}$, completing the proof for \underline{Claim 1} of Theorem~\ref{theorem08}.

\underline{Claim 2.} To analyze the monotonicity of $EC^{(l,\epsilon)}(\theta_{i})$ with respect to $\overline{\bm{\nu}}^{(1)}_{i}$, we can rewrite Eq.~(\ref{equation03}) as follows:
\begin{align}\label{equation14}
	&n\log_{2}\!\left(1\!+\!\overline{\bm{\nu}}^{(1)}_{i}\gamma^{(l)}_{i}\right)\!-\!\sqrt{\!n\!\left(1\!-\!\frac{1}{(1\!+\!\overline{\bm{\nu}}^{(1)}_{i}\gamma^{(l)}_{i})^{2}}\right)} \!\left[\frac{Q^{-1}\!(\epsilon_{i})}{\log 2}\right]\!
	\nonumber\\
	&\quad -\!\log R^{(l)}_{i} =0.
\end{align}
Observing Eq.~(\ref{equation14}), we note that it represents a quadratic function with respect to $n$. 
Consequently, for $\epsilon_{i}\in(0,1/2)$, the positive root of Eq.~(\ref{equation14}) can be calculated as follows:
\begin{align}\label{equation15}
	\sqrt{n}=&\,\frac{1}{2\log_{2}\left(1+\overline{\bm{\nu}}^{(1)}_{i}\gamma^{(l)}_{i}\right)} \Bigg(\frac{Q^{-1}(\epsilon_{i})\sqrt{\overline{\bm{\nu}}^{(1)}_{i}\gamma^{(l)}_{i}(\overline{\bm{\nu}}^{(1)}_{i}\gamma^{(l)}_{i}\!+\!2)}}{\log2(1+\overline{\bm{\nu}}^{(1)}_{i}\gamma^{(l)}_{i})}
	\nonumber\\
	&+\!\sqrt{\frac{\left(Q^{-1}(\epsilon_{i})\right)^{2}\left[\overline{\bm{\nu}}^{(1)}_{i}\gamma^{(l)}_{i}(\overline{\bm{\nu}}^{(1)}_{i}\gamma^{(l)}_{i}\!+\!2)\right]}{2\log2(1+\overline{\bm{\nu}}^{(1)}_{i}\gamma^{(l)}_{i})^{2}}}+4\log R^{(l)}_{i}\!\Bigg)\nonumber \\
	>&\,\frac{Q^{-1}(\epsilon_{i})\sqrt{\overline{\bm{\nu}}^{(1)}_{i}\gamma^{(l)}_{i}(\overline{\bm{\nu}}^{(1)}_{i}\gamma^{(l)}_{i}+2)}}{\left(1+\overline{\bm{\nu}}^{(1)}_{i}\gamma^{(l)}_{i}\right)\log(1+\overline{\bm{\nu}}^{(1)}_{i}\gamma^{(l)}_{i})}>0.
\end{align}
Based on the results in \underline{Claim 1}, the first-order derivative of $\log R^{(l)}_{i}$ in terms of $\overline{\bm{\nu}}^{(1)}_{i}$ is derived as follows:
\begin{align}\label{equation122}
	\frac{\partial \log R^{(l)}_{i}}{\partial \overline{\bm{\nu}}^{(1)}_{i}}\!=&\,\frac{n\gamma^{(l)}_{i}}{\log2\left(1\!+\!\overline{\bm{\nu}}^{(1)}_{i}\gamma^{(l)}_{i}\right)}\!-\!{\gamma^{(l)}_{i}}\Bigg\{\left(1+\overline{\bm{\nu}}^{(1)}_{i}\gamma^{(l)}_{i}\right)^{2}
		\nonumber\\
	&\times\sqrt{\overline{\bm{\nu}}^{(1)}_{i}\gamma^{(l)}_{i}(\overline{\bm{\nu}}^{(1)}_{i}\gamma^{(l)}_{i}\!+\!2)}\Bigg\}^{-1}  \frac{\sqrt{n}Q^{-1}(\epsilon_{i})}{\log 2} \nonumber \\
	=&\,\frac{\sqrt{n}\gamma^{(l)}_{i}}{\log2\left(1+\overline{\bm{\nu}}^{(1)}_{i}\gamma^{(l)}_{i}\right)}\vast[\sqrt{n}-Q^{-1}(\epsilon_{i}) 
	\nonumber\\
	&\qquad \times\! \frac{1}{\left(1\!+\!\overline{\bm{\nu}}^{(1)}_{i}\gamma^{(l)}_{i}\right)\!\sqrt{\overline{\bm{\nu}}^{(1)}_{i}\gamma^{(l)}_{i}(\overline{\bm{\nu}}^{(1)}_{i}\gamma^{(l)}_{i}\!+\!2)}}\!\vast].
\end{align}
Using Eq.~(\ref{equation15}), we can obtain the following equations:
\begin{align}\label{equation122a}
&	\frac{\partial \log R^{(l)}_{i}}{\partial \overline{\bm{\nu}}^{(1)}_{i}}\nonumber\\
&\quad\!>\!\frac{\sqrt{n}\gamma^{(l)}_{i}}{\log2\left(1\!+\!\overline{\bm{\nu}}^{(1)}_{i}\gamma^{(l)}_{i}\right)}\!\vvast[\!\frac{Q^{-1}(\epsilon_{i})\sqrt{\overline{\bm{\nu}}^{(1)}_{i}\gamma^{(l)}_{i}(\overline{\bm{\nu}}^{(1)}_{i}\gamma^{(l)}_{i}\!+\!2)}}{\left(1+\overline{\bm{\nu}}^{(1)}_{i}\gamma^{(l)}_{i}\right)\log(1\!+\!\overline{\bm{\nu}}^{(1)}_{i}\gamma^{(l)}_{i})} 
	\nonumber\\
	&\qquad\!-\!Q^{-1}(\epsilon_{i})  \frac{1}{\left(1\!+\!\overline{\bm{\nu}}^{(1)}_{i}\gamma^{(l)}_{i}\right)\sqrt{\overline{\bm{\nu}}^{(1)}_{i}\gamma^{(l)}_{i}(\overline{\bm{\nu}}^{(1)}_{i}\gamma^{(l)}_{i}\!+\!2)}}\vvast]\nonumber\\
	&\quad\!=\!\frac{\sqrt{n}Q^{-1}(\epsilon_{i})\gamma^{(l)}_{i}}{\log2\left(1+\overline{\bm{\nu}}^{(1)}_{i}\gamma^{(l)}_{i}\right)}\vast[\frac{\sqrt{\overline{\bm{\nu}}^{(1)}_{i}\gamma^{(l)}_{i}(\overline{\bm{\nu}}^{(1)}_{i}\gamma^{(l)}_{i}+2)}}{\left(1+\overline{\bm{\nu}}^{(1)}_{i}\gamma^{(l)}_{i}\right)\log(1+\overline{\bm{\nu}}^{(1)}_{i}\gamma^{(l)}_{i})} 	\nonumber\\
	&\qquad\!-\!\frac{1}{\left(1+\overline{\bm{\nu}}^{(1)}_{i}\gamma^{(l)}_{i}\right)\sqrt{\overline{\bm{\nu}}^{(1)}_{i}\gamma^{(l)}_{i}(\overline{\bm{\nu}}^{(1)}_{i}\gamma^{(l)}_{i}+2)}}\vast] \nonumber\\
	&\qquad \!\times\! \frac{\overline{\bm{\nu}}^{(1)}_{i}\gamma^{(l)}_{i}(\overline{\bm{\nu}}^{(1)}_{i}\gamma^{(l)}_{i}+2)\!-\!\log\left(1\!+\!\overline{\bm{\nu}}^{(1)}_{i}\gamma^{(l)}_{i}\right)}{\left(1\!+\!\overline{\bm{\nu}}^{(1)}_{i}\gamma^{(l)}_{i}\right)\log(1\!+\!\overline{\bm{\nu}}^{(1)}_{i}\gamma^{(l)}_{i})\sqrt{\overline{\bm{\nu}}^{(1)}_{i}\gamma^{(l)}_{i}(\overline{\bm{\nu}}^{(1)}_{i}\gamma^{(l)}_{i}\!+\!2)}} \nonumber\\
	&\qquad \!\times \!\frac{\sqrt{n}Q^{-1}(\epsilon_{i})\gamma^{(l)}_{i}}{\log2\left(1+\overline{\bm{\nu}}^{(1)}_{i}\gamma^{(l)}_{i}\right)} 
	>0.
\end{align}
Consequently, we can obtain 
\begin{align}\label{equation125}
	\frac{\partial EC^{(l,\epsilon)}(\theta_{i})}{\partial \overline{\bm{\nu}}^{(1)}_{i}}=\frac{\partial EC^{(l,\epsilon)}(\theta_{i})}{\partial \log R^{(l)}_{i}}\frac{ \partial \log R^{(l)}_{i}}{\partial \overline{\bm{\nu}}^{(1)}_{i}}>0,
\end{align}
implying that $EC^{(l,\epsilon)}(\theta_{i})$ increases in terms of $\overline{\bm{\nu}}^{(1)}_{i}$ when $\epsilon_{i}\in(0,1/2)$, completing the proof for \underline{Claim 2} of Theorem~\ref{theorem08}.
Therefore, we complete the proof of Theorem~\ref{theorem08}.

\section{Proof for Theorem 5}
Initially, let us compute the first-order derivative of $EE^{(l,\epsilon)}(\theta_{i})$ concerning $\overline{\bm{\nu}}^{(1)}_{i}$, which is expressed as follows:
\begin{align}\label{equation124}
	\frac{\partial EE^{(l,\epsilon)}(\theta_{i})}{\partial \overline{\bm{\nu}}^{(1)}_{i}}=\frac{\frac{\partial EC^{(l,\epsilon)}(\theta_{i})}{\partial \overline{\bm{\nu}}^{(1)}_{i}}\left(\eta_{i}\overline{{\cal P}}_{i}(\bm{\nu})+{\cal P}_{c}\right)-\eta_{i} EC^{(l,\epsilon)}(\theta_{i})}{\left(\eta_{i}\overline{{\cal P}}_{i}(\bm{\nu})+{\cal P}_{c}\right)^{2}}.
\end{align}
We define the auxiliary function $G(\overline{\bm{\nu}}^{(1)}_{i})$ in the following manner:
	\begin{align}\label{equation129}
		G(\overline{\bm{\nu}}^{(1)}_{i})\triangleq\frac{\partial EC^{(l,\epsilon)}(\theta_{i})}{\partial \overline{\bm{\nu}}^{(1)}_{i}}\left(\eta_{i}\overline{{\cal P}}_{i}(\bm{\nu})+{\cal P}_{c}\right)-\eta_{i} EC^{(l,\epsilon)}(\theta_{i}).
	\end{align}
	Taking the first-order derivative of function $G(\overline{\bm{\nu}}^{(1)}_{i})$ with respect to $\overline{\bm{\nu}}^{(1)}_{i}$, we have
	\begin{align}\label{equation30}
		\frac{\partial G(\overline{\bm{\nu}}^{(1)}_{i})}{\partial \overline{\bm{\nu}}^{(1)}_{i}}=\frac{\partial^{2} EC^{(l,\epsilon)}(\theta_{i})}{\partial (\overline{\bm{\nu}}^{(1)}_{i})^{2}}\left(\eta_{i}\overline{{\cal P}}_{i}(\bm{\nu})+{\cal P}_{c}\right).
	\end{align}
	As demonstrated by Theorem~\ref{theorem02}, the concavity of uplink $\epsilon$-effective capacity $EC^{(l,\epsilon)}(\theta_{i})$ in $\overline{\bm{\nu}}^{(1)}_{i}$ is established when $n> \max \left\{\left(n^{\text{rt}}\right)^{2}, n^{\text{th}}\right\}$, we prove $\frac{\partial^{2} EC^{(l,\epsilon)}(\theta_{i})}{\partial (\overline{\bm{\nu}}^{(1)}_{i})^{2}}<0$. 
	Consequently, since $\left(\eta_{i}\overline{{\cal P}}_{i}(\bm{\nu})+{\cal P}_{c}\right)>0$, we establish that $\frac{\partial G(\overline{\bm{\nu}}^{(1)}_{i})}{\partial \overline{\bm{\nu}}^{(1)}_{i}}<0$, indicating that $G(\overline{\bm{\nu}}^{(1)}_{i})$ monotonically decreases with respect to $\overline{\bm{\nu}}^{(1)}_{i}$. 
	Moreover, to derive the optimal solution that maximizes $\mathbf{P_{5}}$, we set $G(\overline{\bm{\nu}}^{(1)}_{i})|_{\overline{\bm{\nu}}^{(1)}_{i}=\overline{\bm{\nu}}_{i}^{(l,\text{opt})}}=0$ using Eq.~(\ref{equation129}).
	Subsequently, the recursive equation is obtained as follows:
	\begin{align}\label{equation31}
		\overline{\bm{\nu}}_{i}^{(l,\text{opt})}=&\,\frac{EC^{(l,\epsilon)}(\theta_{i})}{\frac{\partial EC^{(l,\epsilon)}(\theta_{i})}{\partial \overline{\bm{\nu}}^{(1)}_{i}}\Big|_{\overline{\bm{\nu}}^{(1)}_{i}=\overline{\bm{\nu}}_{i}^{(l,\text{opt})}}}-\frac{{\cal P}_{c}}{\eta_{i}},
	\end{align}
	which can be resolved numerically, and thus	concluding the proof of Theorem~\ref{theorem09}.
\end{appendices}

\nocite{*}
\bibliographystyle{IEEEtran}
\bibliography{IEEEabrv,myref_info}

\end{document}